\documentclass[english,aps,prx,twocolumn,superscriptaddress,longbibliography]{revtex4-2}

\usepackage{amssymb}
\usepackage{natbib}
\usepackage{epstopdf}
\usepackage{braket}
\usepackage{physics}
\usepackage{mathrsfs}
\usepackage{upgreek}
\usepackage{todonotes}
\usepackage[colorlinks=true, allcolors=blue]{hyperref}
\usepackage{lineno}
\usepackage{cancel}
\usepackage{xcolor}
\usepackage{graphicx}
\usepackage{color}
\usepackage{comment}
\usepackage[normalem]{ulem}
\usepackage{amsmath}
\usepackage{lipsum}
\usepackage{siunitx}

\definecolor{mycolor}{RGB}{0, 0, 0}

\begin{document}
\raggedbottom
\title{Non-equilibrium transport reveals energy level degeneracy}
\author{Artem~O.~Denisov}
\email{adenisov@phys.ethz.ch}
\author{Christoph~Adam}
\author{Hadrien~Duprez}
\author{Jessica~Richter}
\author{Zhuoyu~Chen}
\affiliation{Laboratory for Solid State Physics, ETH Zurich, CH-8093 Zurich, Switzerland}
\affiliation{Quantum Center, ETH Zürich, CH-8093 Zürich, Switzerland}
\author{Andrea Hofmann}
\affiliation{University of Basel, Klingelbergstrasse 82, 4056 Basel, Switzerland}
\author{Kenji~Watanabe}
\affiliation{Research Center for Functional Materials, National Institute for Materials Science, 1-1 Namiki, Tsukuba 305-0044, Japan}
\author{Takashi~Taniguchi}
\affiliation{International Center for Materials Nanoarchitectonics,
National Institute for Materials Science, 1-1 Namiki, Tsukuba 305-0044, Japan}
\author{Thomas~Ihn}
\author{Klaus~Ensslin}
\affiliation{Laboratory for Solid State Physics, ETH Zurich, CH-8093 Zurich, Switzerland}
\affiliation{Quantum Center, ETH Zürich, CH-8093 Zürich, Switzerland}



\begin{abstract}
We demonstrate a method to determine energy level degeneracies using non-equilibrium electronic transport through voltage-biased quantum dots. 
We establish the general validity of this approach using single and double quantum dots in bilayer graphene and GaAs. 
Unlike established methods based on entropy measurements or time-resolved tunneling statistics, our approach achieves comparable precision without requiring calibrated electron heating or real-time charge detection.
We resolve the predicted symmetric shell structure in bilayer graphene quantum dots, including a singlet ground state at half filling and the ground state degeneracies of the first 13 carriers. 
Extending the method to double quantum dots, we observe degeneracy doubling associated with bonding and antibonding orbitals for a single carrier and a fourfold degeneracy for two carriers, previously inaccessible with existing techniques.
These results establish a conceptually general and experimentally straightforward approach for probing energy level degeneracies in complex quantum systems.

\end{abstract}


\maketitle

\twocolumngrid

\section{Introduction}

\begin{figure*}[tbh!]
	\includegraphics[width=1.70\columnwidth]{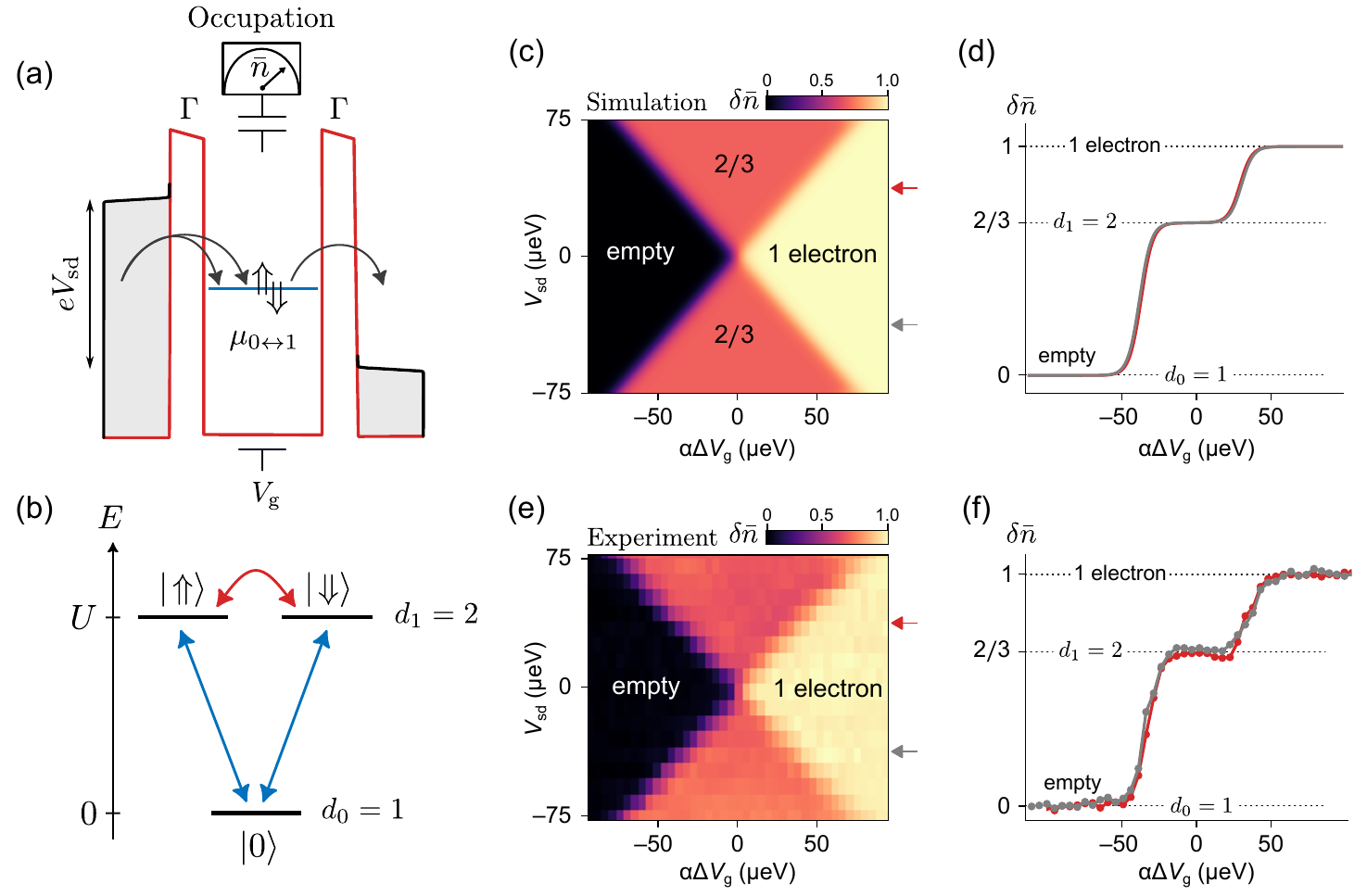}
	\caption{
(a)~Schematic energy diagram of an out-of-equilibrium quantum dot symmetrically coupled to two voltage-biased reservoirs with tunneling rate $\Gamma$. A resonant level within the bias window enables transitions from the empty-dot state $|0\rangle$ to a doubly degenerate spin-$1/2$-like level $|1\rangle$. Owing to this degeneracy, the effective tunneling-in rate from the left reservoir is doubled.
(b)~Two-level energy diagram for the $0$ and $1$ charge states with ground-state degeneracies $d_{0}=1$ and $d_{1}=2$. Blue arrows indicate allowed charge transitions, while the red arrow denotes a spin-flip process. $U$ denotes the single-particle quantization energy.
(c)~Rate-equation simulation of the average dot occupation as a function of plunger gate voltage and bias voltage. Within the bias window, the occupation approaches the universal value $2/3$, reflecting the degeneracy ratio between the final and initial charge states.
(d)~Horizontal line cuts from (c) taken at $V_{\mathrm{sd}} =\SI{\pm35}{\micro\electronvolt}$, as indicated by the red and gray arrows.
(e)~Measured average charge occupation of a bilayer graphene quantum dot as a function of plunger gate voltage and bias voltage. The occupation inside the bias window approaches the same universal value of $2/3$, corresponding to the lowest Kramers doublet acting as an effective spin-$1/2$ state.
(f)~Horizontal line cuts from (e) taken at $V_{\mathrm{sd}} \approx\SI{\pm35}{\micro\electronvolt}$, as indicated by the red and gray arrows.
}

	\label{fig_0}
\end{figure*}

The degeneracy of the ground state offers direct insights into the symmetries of a quantum system~\cite{landau1981quantum}. 
Concepts like topological protection rely fundamentally on a robust ground-state degeneracy~\cite{freedman2003topological}, while contemporary qubit-control strategies operate with degenerate multispin states~\cite{Russ2025}.
An experimentally accessible ground-state degeneracy serves as a direct indicator of symmetry breaking in phases such as superconductors and ferromagnets~\cite{anderson1997basic}, while also providing a powerful means of distinguishing between phases with robust topological invariants, such as quantum Hall states and Chern insulators~\cite{HasanKane}. 
Revealing the ground state degeneracy of Majorana zero modes in quantum dot chains~\cite{10.21468/SciPostPhys.18.6.206} and Moore-Read quasiparticles in fractional quantum Hall states~\cite{PhysRevLett.110.106805} can give an unambiguous signature of the nonlocal nature of the topological state, which cannot be accessed through conventional observables such as conductance.

An intuitive way to probe degeneracy is to lift it using external magnetic or electric fields. However, such perturbations do not always reveal the true multiplicity, and some exotic states couple only weakly or not at all to either field~\cite{kitaev2001unpaired,PhysRevB.61.10267}.
In these cases the relevant tuning parameter can become nontrivial, such as a superconducting phase~\cite{tenHaaf2025probing} or Coulomb exchange~\cite{pecker2013observation}.

One direct way to extract the ground-state degeneracy is to measure thermodynamic entropy of the system~\cite{kuntsevich2015strongly, rozen2021entropic, fedoc2021cascade}, which is linked to the number of available microstates.
We emphasize that, in the context of this paper, all microstates with energy spacings much smaller than the temperature or, where specified, the applied bias voltage are considered degenerate.
Recently, an elegant method for measuring the entropy of only a few particles in quantum dot systems was developed, and it has demonstrated high precision in revealing the degeneracies of one- and two-particle states~\cite{hartman2018direct,PhysRevLett.129.227702,vbbj-138r,gl59-td1w}. 
Its essence lies in the use of Maxwell relations~\cite{e24030417,pyurbeeva2021thermodynamic}, which map the problem onto a tunnelling experiment in which the dot occupation is measured by a charge sensor while the reservoir temperature is modulated. 
Most importantly, there are several straightforward proposals for naturally extending this technique to probe more exotic physics, such as distinguishing topologically non-trivial states~\cite{PhysRevB.92.195312, PhysRevLett.123.147702} and investigating fractional quantum Hall edge states~\cite{PhysRevLett.131.016601}.

The main drawback of Maxwell-relation--based techniques is the difficulty of creating a controlled temperature bias in mesoscopic quantum dot devices at mK temperatures. The heating has to be either extremely slow, when the temperature of the entire system is elevated, or it requires a local galvanic heater~\cite{hartman2018direct} that must be precisely calibrated using some sort of a primary thermometry~\cite{iftikhar2016primary}.
This approach is considerably more involved than a standard transport experiment, in which a known voltage bias is applied across the quantum dot and the resulting current is measured. Nevertheless, both types of measurements ultimately perturb the effective reservoir distribution at the dot’s chemical potential and, in principle, should provide access to the same information about the level degeneracy. 

We therefore developed a technique that is compatible with entropy-based experimental architectures while avoiding heating, at the cost of introducing a second reservoir coupled to the dot. It relies solely on measuring the time-averaged non-equilibrium charge occupation or the transport current through a voltage-biased quantum dot. Additionally, the non-equilibrium method
provides energy resolution (with electronic temperature precision $\sim k_{\mathrm{B}}T$), since all possible transitions inside the bias window are involved. Finally, it does not require real-time resolution of individual tunnelling events or the low tunnelling rates needed in counting experiments~\cite{GUSTAVSSON2009191} that are also used to probe degeneracy~\cite{PhysRevLett.117.206803}. This is particularly important for quantum dot systems realized in a material with intrinsically gapless or small gap spectra, such as multilayer graphene.

\section{General Framework}

Before addressing the more general case, we illustrate the method by extracting the degeneracy $d_{1}=2$ (corresponding to an entropy change $\Delta S = k_{\mathrm{B}}\ln 2$) of the simplest system: a single-carrier spin-$1/2$ level with no orbital degeneracy~\cite{hartman2018direct}.
Figure~\ref{fig_0}(a) shows the schematic experimental setup, in which the average (excess) occupation $\delta\bar{n}$ of a voltage biased quantum dot is monitored using a nearby charge sensor (see the Appendix~\ref{app:fab} for device schematics and fabrication details).
The dot is symmetrically coupled to two reservoirs with intrinsic tunnelling rates~$\Gamma$.
For clarity, we consider the weak-coupling regime in which the master equation is applicable~\cite{ihn2010semiconductor, PhysRevB.65.045317,PhysRevB.44.6199, PhysRevB.44.1646} provided that $\Gamma \ll k_{\mathrm{B}}T \ll eV_{\mathrm{sd}}$, with the electron temperature $T \approx \SI{30}{\milli\kelvin}$ being unchanged throughout the experiment.
In this limit, finite level broadening and Kondo-type correlations can be neglected. Their treatment requires specialized theoretical tools beyond the scope of this work, even though the experimental setup would remain unchanged~\cite{PhysRevLett.70.2601, PhysRevLett.129.227702}.

The energetically allowed transitions lying well inside the bias window are shown in Fig.~\ref{fig_0}(b). The empty dot can be filled in two distinct ways by a single carrier with effective spin-up or spin-down, however, it can be emptied only into a single, non-degenerate empty state. Therefore, the effective spin-conserving tunnelling-in rate is doubled, $\Gamma_{\mathrm{in}}=d_{1}\Gamma=2\Gamma$, while the tunnelling-out rate remains intrinsic $\Gamma_{\mathrm{out}}=\Gamma$~\cite{PhysRevLett.117.206803,duprez2024spin}. 
As a result, the average occupation of the dot when the transition level lies inside the bias window takes the universal value of
\begin{equation}
    \delta \bar{n}
    = \frac{1/\Gamma_{\mathrm{out}}}{1/\Gamma_{\mathrm{out}} + 1/\Gamma_{\mathrm{in}}}
    = \frac{d_{1}}{d_{1}+1}
    = \frac{2}{3}.
\end{equation}

\begin{figure*}[tbh!]
	\includegraphics[width=1.75\columnwidth]{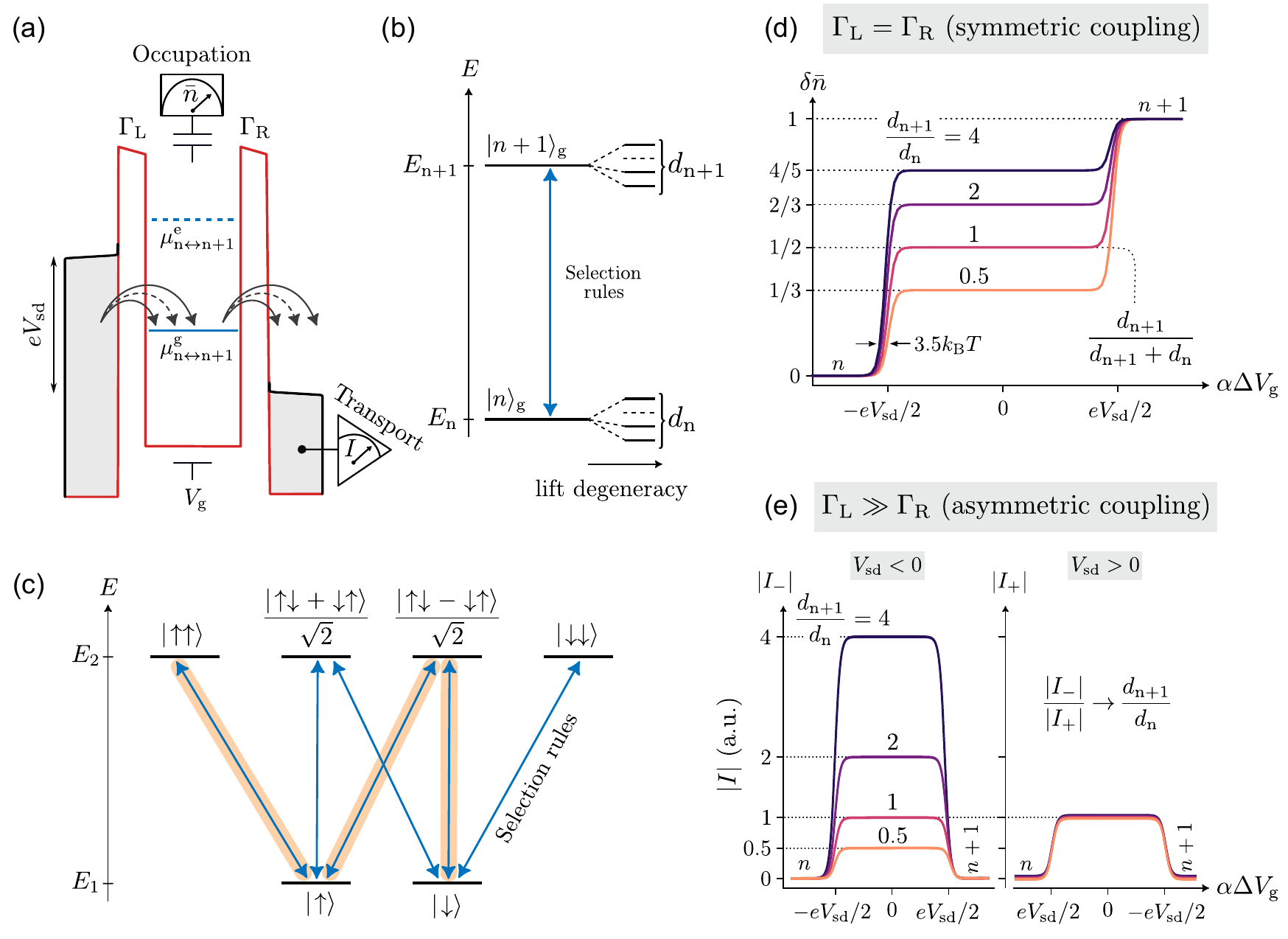}
	\caption{
(a)~Schematic of an out-of-equilibrium quantum dot coupled to two voltage-biased reservoirs with tunneling rates $\Gamma_{\mathrm{L}}$ and $\Gamma_{\mathrm{R}}$. A resonant level within the bias window enables transitions from the $|n\rangle$ ground state to the ground (solid) and excited (dashed) $|n\!+\!1\rangle$ charge states. The average dot occupation $\bar{n}$ is monitored by a nearby charge sensor, while the transport current $I$ is measured at the drain.
(b)~Two-level energy diagram for the $n$ and $n\!+\!1$ charge states with ground-state degeneracies $d_{\mathrm{n}}$ and $d_{\mathrm{n+1}}$.
(c)~Transitions between degenerate manifolds of one and two spin$-1/2$ particles energy states allowed by selection rules (blue arrows). The orange path illustrates an indirect transition between states that are not directly coupled.
(d)~Symmetric coupling, $\Gamma_{\mathrm{L}}=\Gamma_{\mathrm{R}}$. Average dot occupation as a function of plunger gate voltage $\alpha \Delta V_{\mathrm{g}}$ for different degeneracy ratios $d_{\mathrm{n+1}}/d_{\mathrm{n}}$, where $\alpha$ is the lever arm. The occupation plateau within the bias window directly reflects this ratio.
(e)~Strongly asymmetric coupling, $\Gamma_{\mathrm{L}}\gg\Gamma_{\mathrm{R}}$. Absolute transport current at positive ($|I_{+}|$) and negative ($|I_{-}|$) bias as a function of plunger gate voltage $\alpha eV_{\mathrm{g}}$ for different degeneracy ratios. In this regime, the ratio of current plateaus inside the bias window directly yields $d_{\mathrm{n+1}}/d_{\mathrm{n}}$.
}
	\label{fig_1}
\end{figure*}

Figures~\ref{fig_0}(c,d) show rate-equation simulations of $\delta \bar{n}$ as a function of source–drain bias $V_{\mathrm{sd}}$ and plunger gate voltage $V_{\mathrm{g}}$, which controls the dot chemical potential relative to the leads. The resulting $2/3$ occupation plateau is experimentally confirmed in a bilayer graphene quantum dot for $|eV_{\mathrm{sd}}|\gg k_{\mathrm{B}}T$, as shown in Figs.~\ref{fig_0}(e,f). In the plot of the measurement results, we set the occupations inside the Coulomb-blockaded regions measured by a charge detector to 0 and 1, respectively, such that the height of the step inside the bias window can be directly related to the degeneracy. 

In our experiment, the effective spin-$1/2$ system is realized in a single-carrier bilayer graphene (BLG) quantum dot, whose ground state is a Kramers doublet formed by the lowest states with opposite spin and valley quantum numbers~($|\!\Downarrow\rangle ~=~|K^{-}\!\downarrow\rangle$, $|\!\Uparrow\rangle~=~|K^{+}\!\uparrow\rangle$)~\cite{duprez2024spin,denisov2025spin}. Spin-flip-like processes within the degenerate manifold, indicated by the red arrow in Fig.~\ref{fig_0}(b), do not affect the result provided that the tunnelling matrix elements, and therefore the corresponding rates, are identical for all degenerate states.

\subsection{Charge detection}

We now continue with the more general case shown in Fig.~\ref{fig_1}. 
We consider a quantum dot coupled to the left and right leads with arbitrary intrinsic tunneling rates $\Gamma_{\mathrm{L}}$ and $\Gamma_{\mathrm{R}}$ respectively as sketched in Fig.~\ref{fig_1}(a). 

At this point, we introduce a crucial assumption (see Appendix~\ref{app:framework} for details): for each given state within a degenerate manifold, the orbital part of the wavefunction has the same spatial profile in the vicinity of the left and right tunnel barriers, even though the wavefunctions of different degenerate states may differ and the barrier transparencies themselves may be unequal.
This condition can be realized, for example, if the quantum dot exhibits approximate elliptical symmetry with respect to the left and right tunnel barriers or, ideally, if the two reservoirs are brought sufficiently close together such that they probe the same wavefunction tail of the dot. Under this assumption, the two biased reservoirs can effectively be treated as a single reservoir with a non-equilibrium double-step distribution function~\cite{PhysRevLett.79.3490, PhysRevLett.102.036804,Altimiras2010, PhysRevB.102.085417}. Remarkably, detailed balance between the quantum dot and such a non-equilibrium reservoir can still be established (see Appendix~\ref{app:framework}), which forms a central foundation of the present work.

Simultaneously with the average dot occupation $\delta \bar{n}$, the net DC current $I$ is measured. 
As the electrochemical potential $\mu^{\mathrm{g}}_{n\,\leftrightarrow\,n+1}$ of the QD is tuned into the bias window by the plunger gate, the transitions between the electron ground states $n$ and $n+1$ become allowed, leading to changes in both the dot occupation and the current. 

In the general case, the ground-state manifolds $|n,i\rangle_{\mathrm{g}}$ and $|n+1,j\rangle_{\mathrm{g}}$ comprise sets of orthogonal wavefunctions with degeneracies $d_{n}$ and $d_{n+1}$, respectively, as illustrated in Fig.~\ref{fig_1}(b). 
Remarkably, detailed knowledge of the selection rules and tunneling matrix elements is not necessary for extracting the ratio of degeneracies $d_{n+1}/d_{n}$ or the associated entropy change $\Delta S=k_{\mathrm{B}}\log{({d_{\mathrm{n+1}}}/{d_{\mathrm{n}}})}$. Such rules may be imposed by the conservation of spin, valley, or other quantum numbers during tunnelling~\cite{Garreis2024} from/to the lead.
The only condition we require is that the measurement averaging time greatly exceeds the characteristic tunnelling time, such that thermodynamic equilibrium can be established among all accessible states, that is, detailed balance holds. This equilibrium is ensured by the fact that the resulting state graph (Markov chain) satisfies the Kolmogorov reversibility criterion~\cite{kelly1979reversibility} (a full proof is provided in Appendix~\ref{app:framework}) and is irreducible.

A prominent example illustrating the importance of this condition is shown in Fig.~\ref{fig_1}(c) for the transition between one- and two-electron spin-$1/2$ states (neglecting the exchange energy). The polarized spin-up triplet state $|\!\uparrow\uparrow\rangle$ cannot be reached directly from the spin-down state $|\!\downarrow\rangle$ if spin is conserved during tunnelling. Moreover, the tunnelling matrix elements into polarized and unpolarized triplet states are different. Nevertheless, because the state graph remains irreducible, the polarized triplet states can be accessed through a sequence of intermediate tunnelling processes (highlighted by orange shading), thereby ensuring detailed balance among all connected states.

When detailed balance holds for each degenerate state, the excess dot occupation within the bias window for positive ($+$) and negative ($-$) bias depends on two dimensionless parameters--the degeneracy ratio ${d_{n+1}}/{d_{n}}$ and the tunnelling-rate asymmetry $\Gamma_{\mathrm{L}}/\Gamma_{\mathrm{R}}$--and is given by
\begin{equation}
    \delta \bar{n}_{+}=\frac{d_{\mathrm{n+1}}\Gamma_{\mathrm{L}}}{d_{\mathrm{n+1}}\Gamma_{\mathrm{L}}+d_{\mathrm{n}}\Gamma_{\mathrm{R}}},~~~~
    \delta \bar{n}_{-}=\frac{d_{\mathrm{n+1}}\Gamma_{\mathrm{R}}}{d_{\mathrm{n+1}}\Gamma_{\mathrm{R}}+d_{\mathrm{n}}\Gamma_{\mathrm{L}}}.
    \label{eq7}
\end{equation}

\subsection{Current measurements}

Similarly, the ratio of the currents for opposite bias polarities reads~\cite{PhysRevB.65.045317}
\begin{equation}
    \frac{|I_{-}|}{|I_{+}|}=\frac{d_{\mathrm{n+1}}\Gamma_{\mathrm{L}}+d_{\mathrm{n}}\Gamma_{\mathrm{R}}}{d_{\mathrm{n+1}}\Gamma_{\mathrm{R}}+d_{\mathrm{n}}\Gamma_{\mathrm{L}}}.
    \label{eq8}
\end{equation}

\begin{figure*}[tbh!]
	\includegraphics[width=1.95\columnwidth]{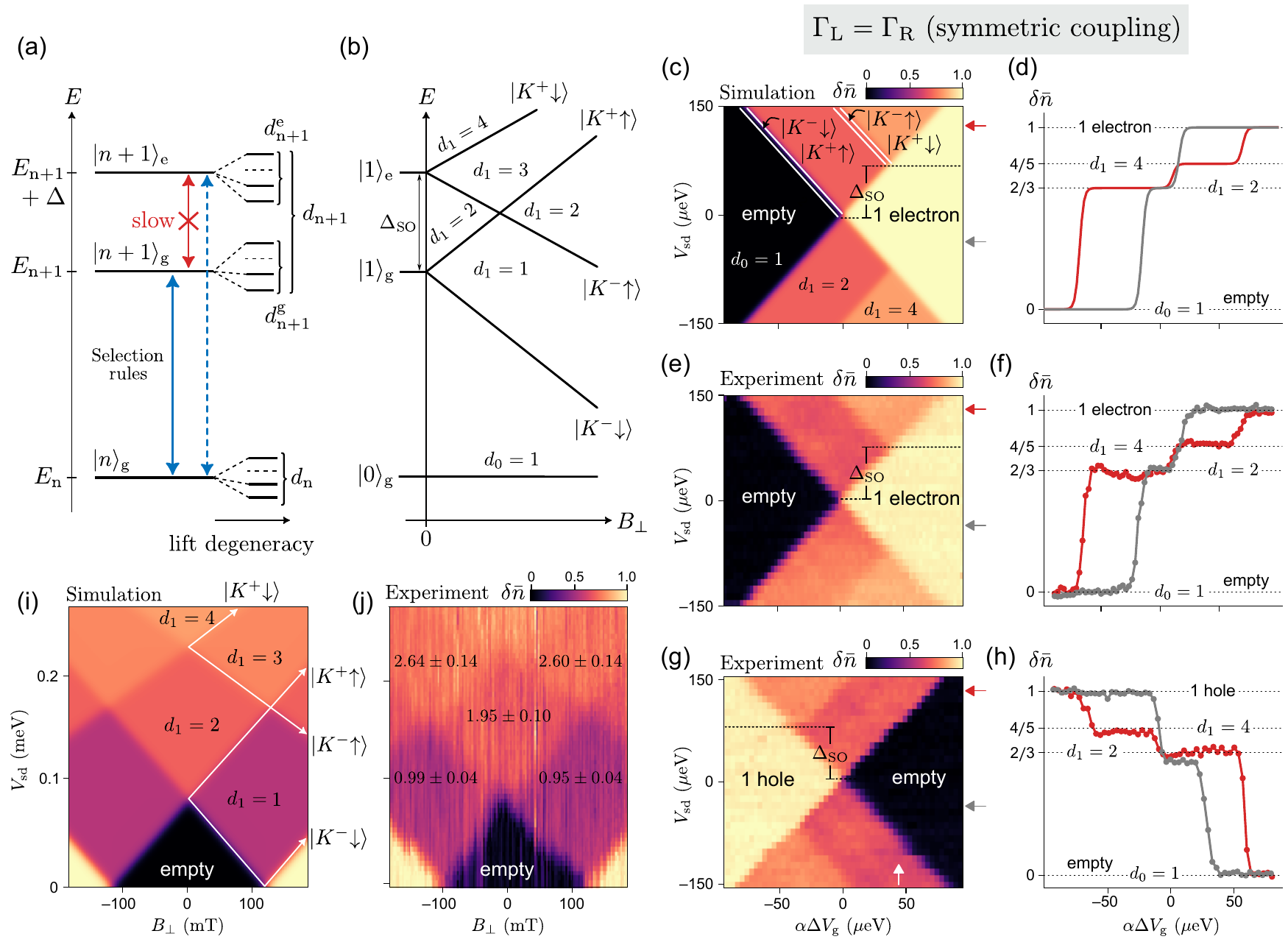}
	\caption{
(a)~Energy diagram of a three-level system for the $n$ and $n\!+\!1$ charge states of the quantum dot, with ground- and excited-state degeneracies $d^{\mathrm{g}}_{\mathrm{n}}$, $d^{\mathrm{g}}_{\mathrm{n+1}}$, and $d^{\mathrm{e}}_{\mathrm{n+1}}$. The excited state of the $n\!+\!1$ configuration is separated from the ground state by an energy $\Delta$. Relaxation between excited and ground states (red arrow) is slow compared with tunneling to and from the leads (blue arrows).
(b)~Single-carrier energy spectrum of a bilayer graphene quantum dot as a function of out-of-plane magnetic field. The two Kramers pairs are separated by the Kane--Mele spin--orbit gap $\Delta_{\mathrm{SO}} \approx \SI{70}{\micro\electronvolt}$. A magnetic field lifts the degeneracy of each pair via spin and valley Zeeman splittings with distinct effective $g$ factors, leading to a stepwise evolution of the effective degeneracy from $d_{1}=1$ to $d_{1}=4$ as more transitions enter the bias window. The empty-dot degeneracy is $d_{0}=1$.
(c)~Rate-equation simulation of the dot occupation as a function of source--drain bias $V_{\mathrm{sd}}$ and plunger gate voltage. The onset of the excited state appears as a step within the bias window as illustrated by the horizontal line cuts in (d). Line-cut positions are indicated by gray and red arrows.
(e,f)~Experimental charge-sensor data corresponding to the simulations in (c,d) for the first-electron transition.
(g,h)~Same as (e,f) for the first-hole transition.The hole spectrum is equivalent to electron up to inversion of the energy axis.
(i,j)~Rate-equation simulation (i) and experimental data (j) of the dot occupation as a function of out-of-plane magnetic field and bias voltage at fixed plunger gate voltage for the first-hole transition, indicated by the white arrow in (g). Solid lines indicate the single-carrier spectrum from (b).
}

	\label{fig_2}
\end{figure*}

We want to emphasize that the absence of effective selection rules or sensitivity to individual matrix elements in the dot dynamics follows from the detailed-balance condition.

The expressions~\eqref{eq7} admit a simple physical interpretation: the occupation corresponds to the fraction of time the dot spends in the $n+1$ charge state, where the effective tunnelling-in or tunnelling-out rates are given by an intrinsic rate multiplied by the degeneracy of the corresponding final $n$ or $n+1$ state.
Similarly, the current~\eqref{eq8} is determined by the characteristic time required for a single electron to tunnel from the left to the right reservoir, 
$I \sim e\left( \frac{1}{d_{\mathrm{n+1}}\Gamma_{\mathrm{L}}} + \frac{1}{d_{\mathrm{n}}\Gamma_{\mathrm{R}}} \right)^{-1}$. 
The pronounced bias asymmetry of the current in Eq.~\eqref{eq8} is striking, as it does not occur in non-interacting conductors under time-reversal symmetry~\cite{BLANTER20001}. 
It originates from strong Coulomb interactions, which prevents tunnelling through the dot when it is occupied (Coulomb blockade) and thereby causes correlations in single electron tunneling \cite{PhysRevB.75.075314}. 
Because the average occupation of the dot depends on the degeneracies and the sign of the bias, the resulting current becomes asymmetric, a behavior that can be derived rigorously from the Meir–Wingreen formalism~\cite{PhysRevLett.68.2512}.

\subsection{Two coupling limits}
The tunnelling rates can typically be tuned close to symmetry, $\Gamma_{\mathrm{L}}/\Gamma_{\mathrm{R}}\simeq 1$, using barrier gates by identifying the point at which the current and/or charge-sensor response becomes symmetric under bias reversal. In this regime and for $\mu^\textnormal{g}_{n\leftrightarrow n+1}$ within the bias window, the measured dot occupation exhibits a plateau at the fractional value
\begin{equation}
\delta \bar{n}=\frac{d_{\mathrm{n+1}}}{d_{\mathrm{n+1}}+d_{\mathrm{n}}},
\label{eq9}
\end{equation}
which directly yields the ratio of degeneracies, as illustrated with the simulated curves in Fig.~\ref{fig_1}(d).

In the opposite, highly asymmetric regime $\Gamma_{\mathrm{L}}/\Gamma_{\mathrm{R}}\gg d_{\mathrm{n+1}}/d_{\mathrm{n}}\sim 1\!-\!10$ (here we exploit the fact that, for most model systems, degeneracies rarely exceed eight), the measured $\delta \bar{n}$ becomes uninformative. It vanishes (or approaches unity) because the electron is either trapped on the dot or the dot remains empty most of the time depending on the bias direction. 
Fortunately, in this regime, the ratio of currents remains well defined and directly yields the ratio of degeneracies from Eq.~\eqref{eq8},
\begin{equation}
\frac{|I_{-}|}{|I_{+}|}=\frac{d_{\mathrm{n+1}}}{d_{\mathrm{n}}}
\label{eq10}.
\end{equation}
as shown in Fig.~\ref{fig_1}(e). 
Reversing the bias changes the absolute value of the transport current because strong Coulomb interactions (blockade) and the resulting correlations in tunneling make the average dot occupation itself bias dependent. This current asymmetry can equivalently be interpreted as a bias-dependent spectral function of the quantum dot~\cite{PhysRevLett.68.2512}, analogous to scanning tunnelling microscopy (STM), where the vacuum gap between STM tip and device forms the slow tunnel barrier~$\Gamma_{\mathrm{R}}$ and the insulator between device and substrate acts as the strongly coupled reservoir~$\Gamma_{\mathrm{L}}$~\cite{Kovarik2024}.

\subsection{Accessing the excited states}

Not only ground-to-ground-state transitions are accessible within our approach. 
In the presence of an additional excited-state manifold $|n+1\rangle_{\mathrm{e}}$ with degeneracy $d^\mathrm{e}_{n+1}$ (case of BLG), separated from the ground state by an energy $\Delta$ as illustrated in Fig.~\ref{fig_2}(a), 
the extracted ratio of degeneracies $d_{\mathrm{n+1}}/d_{\mathrm{n}}$ evolves in a stepwise manner as additional transitions enter the bias window, such that the probed degeneracies add up as $d_{\mathrm{n+1}}=d^{\mathrm{g}}_{\mathrm{n+1}}+d^{\mathrm{e1}}_{\mathrm{n+1}}+\dots$ and $d_{\mathrm{n}}=d^{\mathrm{g}}_{\mathrm{n}}+d^{\mathrm{e1}}_{\mathrm{n}}+\dots$. The presence of excited states serves for a direct validation of the method, as their sequential inclusion in the bias window produces predictable, quantized changes in the measured degeneracy.

In the general case, relaxation (and the corresponding absorption) between excited and ground states must be slow compared with the tunnelling rates to preserve reversibility. This assumption is supported by the long spin and valley relaxation times observed in BLG quantum dots, typically ranging from hundreds of milliseconds to minutes~\cite{Garreis2024, denisov2025spin}, whereas the tunnelling rates in this work lie in the range \SIrange{10}{1}{\mega\hertz}{}. In addition, when all tunnelling matrix elements are equal (as for tunnelling to and from the vacuum state), relaxation processes are irrelevant because the average occupation of the dot does not depend on which particular state participates in the emptying process.

In the sections below, we demonstrate the validity of the expressions in~Eqs.~\eqref{eq7}–\eqref{eq10} for extracting degeneracies using well-known examples of single-carrier quantum dots in bilayer graphene and GaAs (see Appendix~\ref{app:gaas} for the GaAs device). We explore both the symmetric and highly asymmetric coupling regimes, as well as the effect of a magnetic field that lifts the spin and valley degeneracies. We further demonstrate the crossover between these two limits in the intermediate regime $\Gamma_{\mathrm{L}}\sim\Gamma_{\mathrm{R}}$. Finally, we apply the technique to determine the sequence of ground-state degeneracies of many-carrier states in a bilayer graphene quantum dot and to demonstrate, for the first time, the orbital doubling of the degeneracy of the $\frac{1}{\sqrt{2}}\big[(01)\pm(10)\big]$ state in a double quantum dot (DQD).

\section{Symmetric coupling}
We first validate this framework in the situation of symmetric tunnel coupling.
In this section, a bilayer graphene based quantum dot device was intentionally tuned to a barrier-symmetric $\Gamma_{\mathrm{L}}=\Gamma_{\mathrm{R}}$ configuration, ensuring that the extracted degeneracies are not influenced by tunnel-rate asymmetries.

For a single carrier, bilayer graphene quantum dots exhibit spin ($\uparrow,\downarrow$) and valley ($K^{+},K^{-}$) degrees of freedom.
The system provides a well-established example~\cite{denisov2025spin,duprez2024spin,Banszerus2021} of a spectrum containing a doubly degenerate ground state ($|K^{-}\!\downarrow\rangle$, $|K^{+}\!\uparrow\rangle$) and a doubly degenerate excited state ($|K^{-}\!\uparrow\rangle$, $|K^{+}\!\downarrow\rangle$), separated by the spin–orbit coupling gap of typically $\Delta_{\mathrm{SO}} \approx \SI{70}{\micro\electronvolt}$, as sketched in Fig.~\ref{fig_2}(b).

\begin{figure*}[tbh!]
	\includegraphics[width=2\columnwidth]{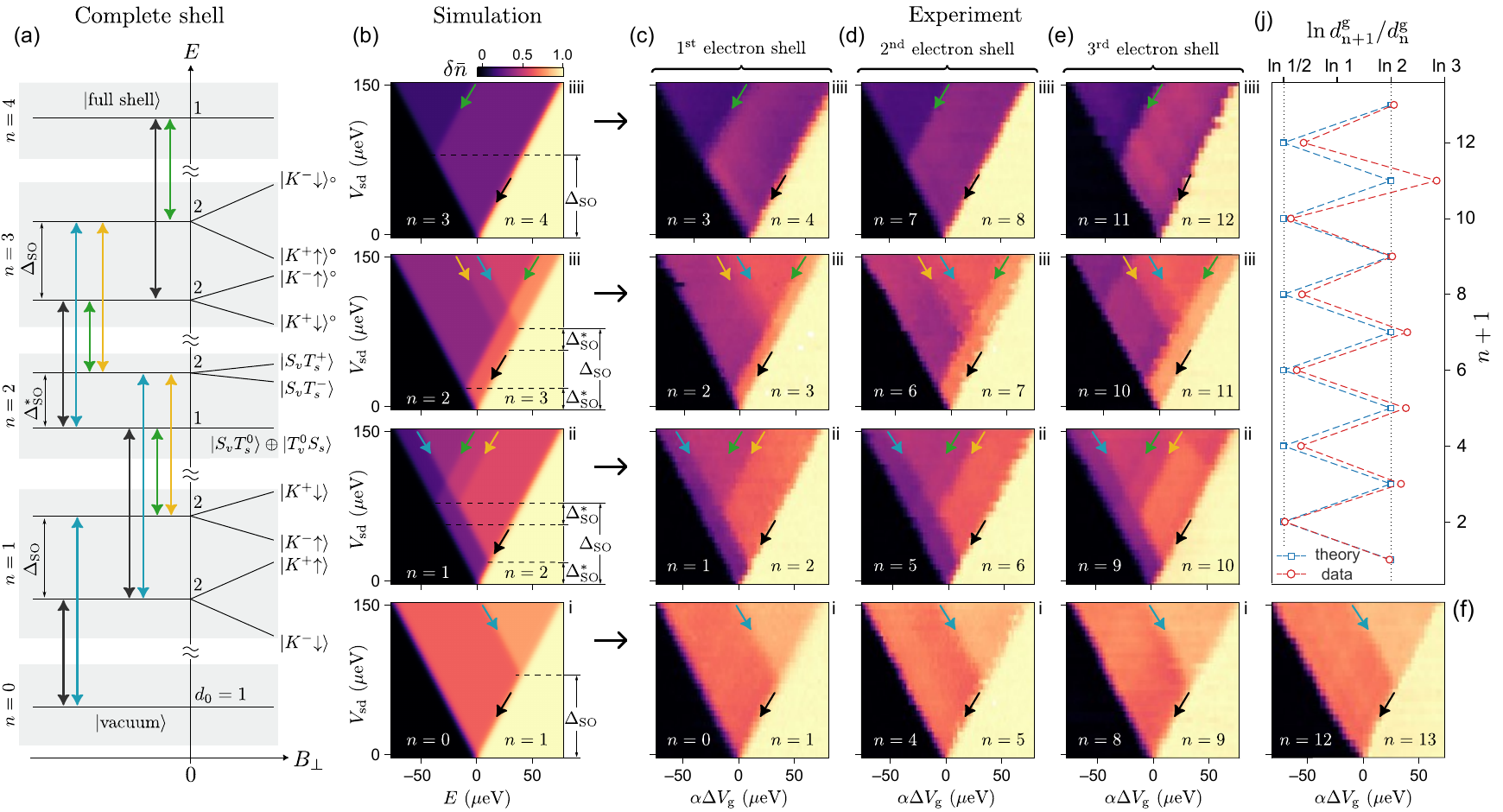}
\caption{
(a)~Energy spectrum of the first electron shell in a bilayer graphene quantum dot as a function of perpendicular magnetic field, illustrating the lifting of degeneracies. Only the ground and first excited states are shown. Colored arrows indicate allowed transitions between the $n$ and $n\!+\!1$ charge states: black (ground--ground), blue (ground--excited), green (excited--ground), and yellow (excited--excited). A full shell of four electrons forms a new effective vacuum state.
(b)~Rate-equation simulation of the quantum-dot occupation for different charge transitions as a function of energy and bias voltage. Arrows indicate the corresponding transitions entering the bias window.
(c--f)~Measured dot occupation as a function of plunger gate voltage (converted to energy) and bias voltage for the first thirteen carriers, grouped into four-electron shells for comparison with the simulations in (b). Colored arrows correspond to the transitions predicted by the model spectrum in (a).
(g)~Extracted ratios of initial and final degeneracies, plotted on a logarithmic scale as a function of carrier number (red circles and dashed line), together with the theoretical values from the spectrum in (a) (blue circles and dashed line).
}

	\label{fig_3}
\end{figure*}

We start the comparison between experiment and theory with the case of zero out-of-plane magnetic field, $B_{\perp}=0$, shown in Figs.~\ref{fig_2}(c--h). 
The dot occupation $\delta \bar{n}$ calculated with the rate equations, is shown in Fig.~\ref{fig_2}(c) as a function of the bias and plunger gate voltage, expressed in energy units. 
With the finite-bias window exceeding the excited state energy, in addition to the ground-to-ground-state transitions, the charge detector further resolves the transition involving the excited state separated from the ground state by $\Delta_{\mathrm{SO}}$. 
The heights of the occupation plateaus of $2/3$ and $4/5$, shown in the horizontal linecuts in Fig.~\ref{fig_2}(d), directly yield degeneracies of $2$ and $4$ for the ground state and for the combined ground-plus-excited-state manifold of the first electron, respectively.
Consistently, the measured data shown in Fig.~\ref{fig_2}(e,f) for the first electron and Fig.~\ref{fig_2}(g,h) for the first hole are closely matched by the corresponding simulations, enabling the extraction of the ground-state degeneracy $d^{\mathrm{g}}_{1}=2$ and the excited-state degeneracy $d^{\mathrm{e}}_{1}=2$, under the assumption of the trivial vacuum degeneracy $d_{0}=1$.
The hole data mirror the electron response upon inversion of the energy axis, consistent with the expected electron–hole symmetry in BLG~\cite{Banszerus2023}, while the weak residual bias dependence observed for the electron dot in Fig.~\ref{fig_2}(e) is attributed to bias-induced gating of the tunnelling barriers.

An applied perpendicular magnetic field lifts the degeneracy of the Kramers pairs with distinct effective $g$ factors determined by their spin and valley indices~\cite{duprez2024spin} as illustrated in Fig.~\ref{fig_2}(b). 
As a result, the measured effective degeneracy of the single-carrier adds up from $d_{1}=1$ to $d_{1}=4$ as additional states enter the bias window. 
The simulated charge occupation in Fig.~\ref{fig_2}(i) exhibits diamond-shaped regions corresponding to a degeneracy $d_{1}=1$ plateau that emerge from the $d_{1}=2$ Kramers doublet plateau as the out-of-plane magnetic field is increased. Similar for the degeneracy $d_{1}=3$ plateau that emerge from the $d_{1}=4$ Kramers doublet plateau at higher bias voltages.
This lifting of the degeneracy is resolved experimentally in Fig.~\ref{fig_2}(j), which shows the measured charge occupation as a function of magnetic field and bias voltage, taken at a fixed plunger gate setting indicated by the white arrow in Fig.~\ref{fig_2}(g).

\section{Shell structure}

Using the method introduced above, we now proceed to extract the ground-state degeneracies at higher carrier numbers in the symmetric tunnel coupling regime. This information is essential for operating spin–valley qubits via exchange interactions~\cite{RevModPhys.95.025003}, as conventional approaches to valley control based on analogues of electron spin resonance~\cite{Laird2013} are ineffective in BLG.

Inside the finite-bias window we resolve the four-electron periodic pattern of excited-state transitions shown in Fig.~\ref{fig_3}(c--f), spanning occupancies from zero to thirteen carriers, where the $0^{\mathrm{th}}$, $4^{\mathrm{th}}$, $8^{\mathrm{th}}$, and $12^{\mathrm{th}}$ electron states effectively act as vacuum states~\cite{Knothe_2022}. 
Each group defines a full four-electron shell. Orbital degeneracies between p-like shells expected for a circular dot are lifted~\cite{PhysRevLett.126.147703}, likely due to elliptic asymmetry of the quantum dot, analogous to observations in carbon nanotubes~\cite{RevModPhys.87.703}. 
Additionally, we confirm that the excited-state spectrum is symmetric about half filling of a given shell, with the spectra for the complementary fillings related by a mirroring of the energy axis and charge occupation. This reflects an intra-shell electron--hole symmetry, distinct from conduction--valence band symmetry in BLG: within a single shell, filling with two electrons is equivalent to filling with two holes, and the one-electron spectrum is the mirror image of the one hole (three electron) spectrum~\cite{Knothe_2022}.

The transition lines observed within the bias window are well described by the theoretical spectrum and the corresponding rate-equation simulations of the occupation shown in Figs.~\ref{fig_3}(a) and (b), respectively. For simplicity, we consider only the ground state and the lowest excited state of each charge carrier number. 
Higher excited states are either absent, as for the first and fourth carriers, or shifted to higher energies by orbital or exchange splittings, as for the second and third carriers~\cite{vbbj-138r,PhysRevLett.127.256802}.

The single-electron spectrum, discussed above, and the three-electron spectrum are identical up to inversion of the energy axis, as confirmed by the measured $0$--$1$, $4$--$5$, $8$--$9$, $12$--$13$ and $3$--$4$, $7$--$8$, $11$--$12$ transitions shown in Fig.~\ref{fig_3}(c--e).
This correspondence arises because the three-electron spectrum is equivalent to the single-hole spectrum within a fully occupied four-electron shell~\cite{PhysRevB.108.125128,Knothe_2022}, as illustrated in Fig.~\ref{fig_3}(a). 
Consistently, the extracted ground-state degeneracies are the same for $d_{1}\simeq d_{5}\simeq d_{9}\simeq d_{13}\simeq 2$ and $d_{3}\simeq d_{7}\simeq d_{11}\simeq 2$, where we assume that the empty dot state degeneracy is $d_{0}=1$. Similarly, the extracted vacuum state degeneracies are close to unity $d_{4}\simeq d_{8}\simeq d_{12}\simeq1$.

The symmetry between the single- and three-electron spectra implies that the transitions to/from half filling ($1$--$2$ and $2$--$3$) are mutually symmetric, while remaining very distinct from the $0$--$1$ and $3$--$4$ transitions.
We found that for all three shells and at each half filling, a low-lying two-carrier excited state is split off by a reduced spin--orbit gap $\Delta^{*}_{\mathrm{SO}} \approx \SI{20}{\micro\electronvolt}$. This gives rise to a pronounced occupation plateau of the ground state, indicating a singlet ground state with degeneracy $d_{2}\simeq d_{6}\simeq d_{10}\simeq 1$.

We summarize the results by plotting the entropy change $\ln(d_{\mathrm{n+1}}/d_{\mathrm{n}})$ for each measured transition in Fig.~\ref{fig_3}(j), in comparison with the theoretical values from Fig.~\ref{fig_3}(a). 
The experimental data follow the expected sequence of ground-state degeneracies $1$--$2$--$1$--$2$--$1$--$\dots$, with the exception of the $10$--$11$ transition, which we attribute to the involvement of higher excited orbital states or impurities.

For a long time, owing to limited spectroscopic resolution, the two-carrier ground state in BLG was believed to be a valley-unpolarized triplet~\cite{PhysRevLett.123.026803, PhysRevB.108.125128} ($|S_{v}T^{+}_{s}\rangle$, $|S_{v}T^{0}_{s}\rangle$, $|S_{v}T^{-}_{s}\rangle$), until this interpretation was revised more recently by entropy-based measurements~\cite{vbbj-138r}. However, the finding was based on a limited data set owing to the experimental complexity of the thermodynamic measurements. 

The revised two carrier picture considers the mixing of two fully spin--valley unpolarized states ($|T^{0}_{v}S_{s}\rangle$ and $|S_{v}T^{0}_{s}\rangle$), which are split by a Coulomb-exchange interaction $J$ that is typically much larger than the bare spin--orbit gap $\Delta_{\mathrm{SO}}$~\cite{vbbj-138r}. The mixing is caused by the same spin--orbit coupling, resulting in a fully unpolarized non-degenerate ground state lowered in energy by $-\Delta^{*}_{\mathrm{SO}}\simeq -\Delta_{\mathrm{SO}}^{2}/J$. 
Although $\Delta^{*}_{\mathrm{SO}}$ in BLG lies close to the resolution limit of typical transport experiments ($\sim\SI{10}{\micro\electronvolt}$), the same mechanism is well established in carbon nanotubes, where substantially larger spin--orbit coupling enables a clear spectroscopic resolution of the corresponding singlet ground-state splitting for the two-carriers state~\cite{pecker2013observation, RevModPhys.87.703, Cleuziou2013}.

\section{Near-symmetric coupling}

\begin{figure*}[tbh!]
	\includegraphics[width=1.4\columnwidth]{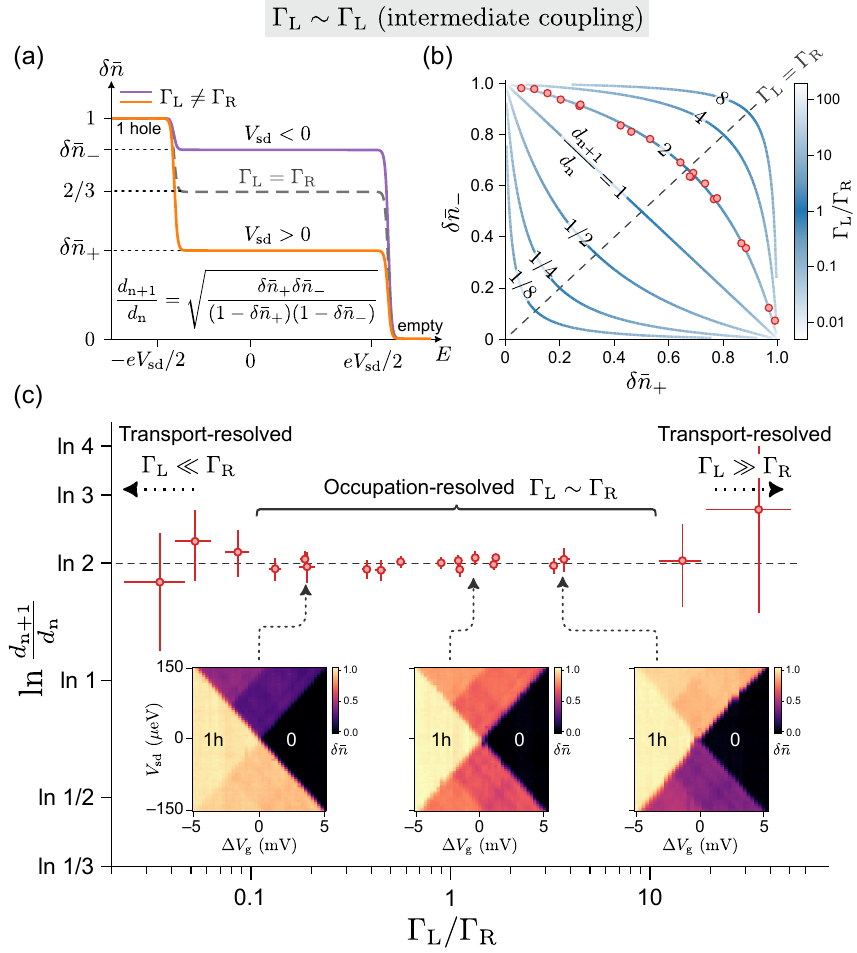}
	\caption{
(a)~Theoretical dot occupation for the first-hole transition as a function of energy in the intermediate-coupling regime, shown for positive (purple) and negative (orange) bias. The dashed line indicates the symmetric-coupling case. The ratio of initial and final degeneracies is extracted from the plateau values $\delta\bar{n}_{+}$ and $\delta\bar{n}_{-}$ inside the bias window.
(b)~Hight of occupation plateaus at positive and negative bias plotted against each other for several degeneracy ratios $d_{\mathrm{n+1}}/d_{\mathrm{n}}$ and coupling asymmetries. Experimental data for the first hole (red circles) fall on the theoretical curve for $d_{\mathrm{n+1}}/d_{\mathrm{n}}=2$. The dashed line marks the symmetric-coupling case.
(c)~Extracted values of $\ln(d_{\mathrm{n+1}}/d_{\mathrm{n}})$ (entropy change) as a function of tunnel-coupling asymmetry. Insets show the measured first-hole transition as a function of bias voltage and plunger gate voltage for the corresponding asymmetries indicated by dashed arrows.
}

	\label{fig_4}
\end{figure*}

So far, we have focused on the symmetric coupling regime, which requires fine tuning of the tunnelling barriers. 
Our technique, however, can also be applied when the intrinsic tunnelling rates are not equal, provided that their asymmetry is not too large. 
In this intermediate regime, $\Gamma_{\mathrm{L}}\sim\Gamma_{\mathrm{R}}$, the charge-sensor response described by Eq.~\eqref{eq7} becomes dependent on the bias direction, yielding distinct occupations $\delta\bar{n}_{-}$ and $\delta\bar{n}_{+}$ for negative and positive bias, respectively, as shown in Fig.~\ref{fig_4}(a). 
These two measured dimensionless quantities $\delta\bar{n}_{\pm}$ provide two independent equations for the two unknown dimensionless parameters, namely the degeneracy ratio $d_{\mathrm{n+1}}/d_{\mathrm{n}}$ and the tunnelling-rate asymmetry $\Gamma_{\mathrm{L}}/\Gamma_{\mathrm{R}}$, which can be extracted as

\begin{align}
 \frac{d_{n+1}}{d_{n}} &= \sqrt{\frac{\delta\bar{n}_{+}\,\delta\bar{n}_{-}}{(1-\delta\bar{n}_{+})(1-\delta\bar{n}_{-})}}, \label{eq11} \\
 \frac{\Gamma_{\mathrm{L}}}{\Gamma_{\mathrm{R}}} &= \sqrt{\frac{\delta\bar{n}_{+}(1-\delta\bar{n}_{-})}{\delta\bar{n}_{-}(1-\delta\bar{n}_{+})}} .
 \label{eq12}
\end{align}
We establish the validity of these relations by analyzing the first-hole transition from Fig.~\ref{fig_2}(g) while changing the asymmetry between the tunnelling barriers. 

Figure~\ref{fig_4}(b) shows the measured occupation plateau height $\delta\bar{n}_{+}$ plotted against its value upon bias reversal, $\delta\bar{n}_{-}$, as the asymmetry between the left and right tunnelling barriers is varied over several orders of magnitude. 
The data for the first-hole transition closely follow the theoretical prediction~\eqref{eq11} for $d_{n+1}/d_{n}=2$ as expected.
The extracted ratio $d_{n+1}/d_{n}$ for each individual data point, including the associated uncertainty, is shown in Fig.~\ref{fig_4}(c). 
The charge-occupation signal provides reliable access to the degeneracy over a broad range of coupling asymmetries, approximately $0.1\lesssim\Gamma_{\mathrm{L}}/\Gamma_{\mathrm{R}}\lesssim10$. Beyond this regime, the uncertainty grows rapidly, as predicted by Eq.~\eqref{eq11}, because the sensor response within the bias window saturates toward $0$ or $1$, as shown in the insets.

We additionally investigate (see Appendix~\ref{app:strongc}) the dependence of the extracted degeneracy on the absolute tunnel-coupling strength $(\Gamma_{\mathrm{L}}+\Gamma_{\mathrm{R}})$ to the leads and observe no deviation from the expected value of $2$, even away from the weak-coupling regime.

Given the robustness of the method based on Eq.~\eqref{eq11}, fine tuning of the tunnel barriers to the symmetric regime is not required to probe degeneracy. A residual tunnel-rate asymmetry in the measured data can be compensated by defining an effective symmetrized occupation,
\begin{equation}
\delta\bar{n}^{\mathrm{sym}} = \dfrac{\sqrt{\delta\bar{n}_{+}\delta\bar{n}_{-}}}{\sqrt{(1-\delta\bar{n}_{+})(1-\delta\bar{n}_{-})}+\sqrt{\delta\bar{n}_{+}\delta\bar{n}_{-}}} ,
\label{eq13}
\end{equation}
such that the resulting plateau heights correspond to those expected for symmetric coupling in Eq.~\eqref{eq9}.

In the strongly asymmetric regime, $\Gamma_{\mathrm{L}}/\Gamma_{\mathrm{R}}\lesssim10^{-2}$ or $\Gamma_{\mathrm{L}}/\Gamma_{\mathrm{R}}\gtrsim10^{2}$, the charge-sensor signal becomes uninformative, while the ratio of transport currents directly yields the degeneracy ratio $d_{n+1}/d_{n}$ in accordance with Eq.~\eqref{eq10}. 
In the following section, we therefore turn to the transport-resolved regime, in which degeneracies can be extracted without the need for symmetric tunnel barriers or a charge sensor.

\section{Asymmetric coupling}

\begin{figure*}[tbh!]
	\includegraphics[width=2\columnwidth]{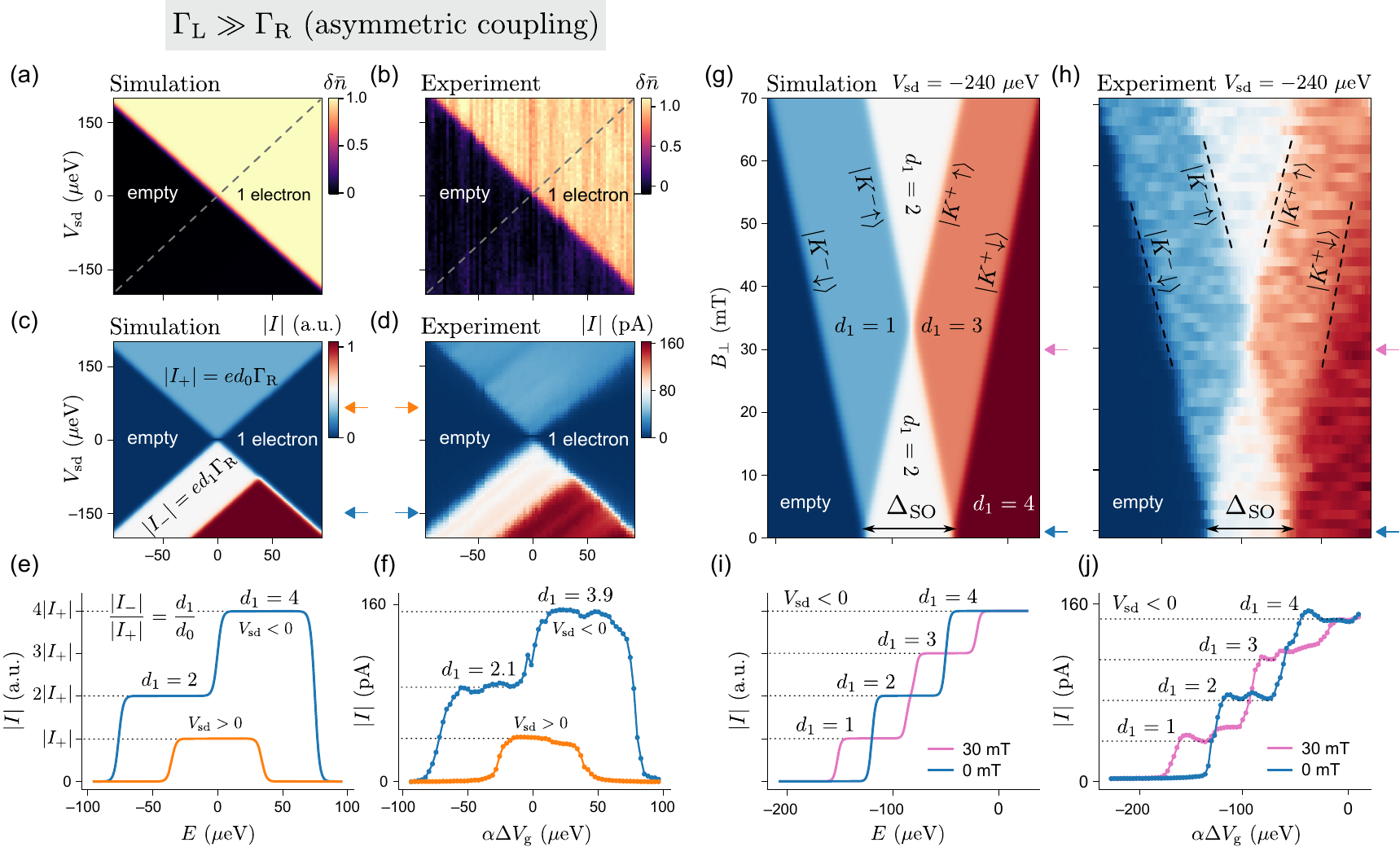}
\caption{
(a,b)~Rate-equation simulations and measured quantum-dot occupation around the $0\!\rightarrow\!1$ electron transition for a dot highly asymmetrically coupled to one lead. Dashed gray lines indicate the edge of bias window.
(c,d)~Corresponding rate-equation simulations and measured transport current. The pronounced bias asymmetry of the current arises from the nontrivial degeneracy of the one-electron state.
(e,f)~Horizontal line cuts of the simulated~(c) and measured~(d) current, respectively, taken at $V_{\mathrm{sd}} = \SI{-150}{\micro\electronvolt}$ (blue arrow) and $\SI{75}{\micro\electronvolt}$ (orange arrow). Dashed lines mark the current plateaus, whose ratio reflects the ground- and excited-state degeneracies of the first carrier, $d^{\mathrm{g}}_{1}=2$ and $d^{\mathrm{e}}_{1}=4$.
(g,h)~Simulated and measured transport current at fixed negative bias $V_{\mathrm{sd}} = \SI{-240}{\micro\electronvolt}$ as a function of energy offset and out-of-plane magnetic field $B_{\perp}$. At zero field, the two Kramers-pair plateaus split into two steps each as spin and valley degeneracies are lifted.
(i,j)~Horizontal current line cuts from (g) and (h) along the energy axis, taken at $B_{\perp} = \SI{0}{\milli\tesla}$ (blue arrow) and $\SI{30}{\milli\tesla}$ (purple arrow) near the crossing point. The initial plateau corresponding to $d_{1}=2$ splits into two equal plateaus, each associated with the addition of a singlet state to the bias window.
}

	\label{fig_5}
\end{figure*}

Figure~\ref{fig_5} illustrates a single-electron transition in the strongly asymmetric coupling regime, $\Gamma_{\mathrm{L}}\gg\Gamma_{\mathrm{R}}$, together with the corresponding rate-equation simulations. 
In this limit, the charge-sensor signal shown in Fig.~\ref{fig_5}(a,b) becomes ineffective, as confirmed by both rate-equation simulations and experiment.
In contrast, the transport current remains highly sensitive to the internal degeneracies, as can be seen from the simplified current expressions~\cite{PhysRevB.65.045317}: 

\begin{align}
  |I_{+}| &= e\,\frac{d_{n+1}\Gamma_{\mathrm{L}}\,d_{n}\Gamma_{\mathrm{R}}}{d_{n+1}\Gamma_{\mathrm{L}}+d_{n}\Gamma_{\mathrm{R}}}
  \;\xrightarrow{\Gamma_{\mathrm{L}}\gg\Gamma_{\mathrm{R}}}\; e\,d_{n}\Gamma_{\mathrm{R}},\label{eq14} \\
  |I_{-}| &= e\,\frac{d_{n+1}\Gamma_{\mathrm{R}}\,d_{n}\Gamma_{\mathrm{L}}}{d_{n+1}\Gamma_{\mathrm{R}}+d_{n}\Gamma_{\mathrm{L}}}
  \;\xrightarrow{\Gamma_{\mathrm{L}}\gg\Gamma_{\mathrm{R}}}\; e\,d_{n+1}\Gamma_{\mathrm{R}}.
  \label{eq15}
\end{align}
These limits are justified when the coupling asymmetry satisfies $\Gamma_{\mathrm{L}}/\Gamma_{\mathrm{R}}\gg d_{n+1}/d_{n}\sim1$–$10$. In the systems studied here, $d_{n+1}/d_{n}$ does not exceed 8, and more generally, degeneracies larger than this are uncommon in experimentally accessible few-electron systems.

An important conclusion from Eqs.~\eqref{eq14} and~\eqref{eq15} is that the current measured at negative bias is sensitive to the effective degeneracy $d_{n+1}$ and therefore probes the excited-state structure of the $n+1$ charge configuration, whereas the current at positive bias reflects the total degeneracy $d_{n}$ and evolves stepwise as additional excited states of the $n$ charge state enter the bias window. 
This leads to an asymmetric current response, shown in Fig.~\ref{fig_5}(c,d), in which the excited state of the first electron in BLG is resolved only at negative bias voltages where $d_{n+1}$ increases from $2$ to $4$. The ratios of the corresponding current plateaus shown in Fig.~\ref{fig_5}(f) yield $d^{\mathrm{g}}_{1}/d_{0}\approx2.1$ and $(d^{\mathrm{g}}_{1}+d^{\mathrm{e}}_{1})/d_{0}\approx3.9$, consistent with the simulation in Fig.~\ref{fig_5}(c,e).

We further validate the technique by lifting the degeneracy with a perpendicular magnetic field, as shown in Fig.~\ref{fig_5}(h), together with the corresponding simulations in Fig.~\ref{fig_5}(g). 
Here, the transport current is measured as a function of plunger gate voltage and perpendicular magnetic field at a fixed bias voltage $V_{\mathrm{sd}} = \SI{-240}{\micro\electronvolt}$. 
The Kramers-doublet plateaus split into distinct branches with degeneracy $d_{1}=1$, analogous to the behavior shown in Fig.~\ref{fig_1}(i,j), resulting in emergence of regions with all possible effective degeneracies from $d_{1}=1,2,3,4$ depending on how many transitions fall within the bias window. 
This splitting into four equivalent current steps at negative bias is highlighted by the line cuts shown in Fig.~\ref{fig_5}(i,j), taken at magnetic fields of $0$ and $\SI{30}{\milli\tesla}$, close to the crossing of two states with identical spin but opposite valley indices.

Taken together, charge sensing in the symmetric and nearly symmetric coupling regimes, along with transport measurements in the strongly asymmetric regime, allow access to degeneracy across the full range of tunnel-coupling asymmetries. 
Importantly, the transport-resolved approach enables the measurement of degeneracy (and entropy) within a minimal experimental framework, requiring neither fine tuning of tunnel barriers, an integrated charge sensor, nor modulated electron temperatures. The only requirement is that the current remains measurable in the regime $\Gamma_{\mathrm{R}} \ll k_{\mathrm{B}}T \sim \SI{1}{\giga\hertz}$, corresponding to currents on the order of $\SIrange{1}{10}{\pico\ampere}$.

\section{Double quantum dot}

\begin{figure*}[tbh!]
	\includegraphics[width=1.8\columnwidth]{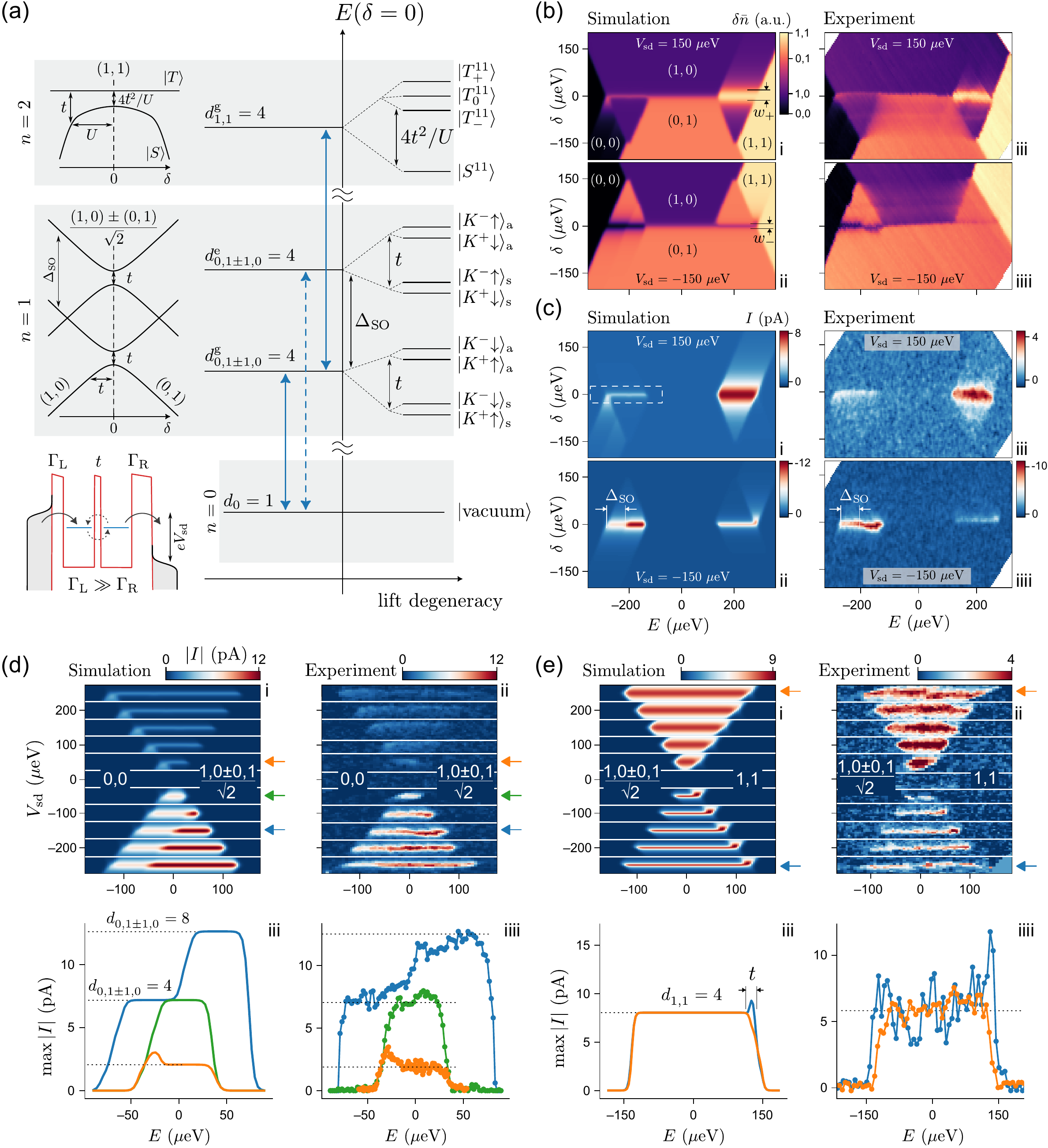}
	\caption{
(a)~Energy spectrum of one- and two-electron states of a bilayer graphene double quantum dot at zero detuning $\delta=0$, with interdot tunnel coupling $t$. For two electrons, only the ground-state manifold is shown, consisting of a Kramers singlet and three triplet states split by the exchange interaction $J=4t^{2}/U$, where $U$ is the charging energy. Colored arrows indicate allowed transitions between the $n$ and $n\!+\!1$ charge states.
(b)~Simulated (i,ii) and measured (iii,iiii) double-dot occupation charge stability diagram for positive ($V_{\mathrm{sd}} = \SI{150}{\micro\electronvolt}$) and negative ($V_{\mathrm{sd}} = \SI{-150}{\micro\electronvolt}$) bias. The widths of the zero-detuning features at positive ($w_{+}$) and negative ($w_{-}$) bias are set by the tunnel coupling $t$ and the lead asymmetry $\Gamma_{\mathrm{L}}/\Gamma_{\mathrm{R}}$.
(c)~Rate-equation simulations (i,ii) and measured transport current (iii,iiii) in the same regions as in (b). The single-carrier excited state is split from the ground state by the spin--orbit gap $\Delta_{\mathrm{SO}} \approx \SI{75}{\micro\electronvolt}$. The dashed box marks the region analyzed in (d,e).
(d)~Simulated (i–iii) and measured (ii–iiii) current near zero detuning for the $(0,0)\rightarrow[(1,0)\pm(0,1)]/\sqrt{2}$ transition, compiled for different bias voltages. Panels (iii) and (iiii) show the simulated and measured maximum current across detuning for $V_{\mathrm{sd}} = \SI{60}{\micro\electronvolt}$ (green), $\SI{-60}{\micro\electronvolt}$ (orange), and $\SI{-150}{\micro\electronvolt}$ (blue). Dashed lines indicate the expected plateau heights for effective degeneracies $d_{10\pm01}=4$ and $d_{10\pm01}=8$.
(e)~Same analysis as in (d) for the $[(1,0)\pm(0,1)]/\sqrt{2}\rightarrow(1,1)$ transition, for which the current shows no pronounced bias asymmetry, consistent with equal degeneracies of the initial and final states.
}

	\label{fig_6}
\end{figure*}

We further extend the technique to the more complex system of a double quantum dot in BLG. 
At zero detuning between two single-carrier charge states of the two coupled dots (0,1) and (1,0), the single-carrier degeneracy doubles $d_{{0,1\pm1,0}}=2d_{1}$ owing to the formation of bonding and antibonding molecular orbitals~\cite{RevModPhys.75.1}. 
This doubling has not previously been accessed through entropy measurements~\cite{gl59-td1w}, primarily because such approaches require elaborate heating protocols and the integration of a second charge sensor. 

The energy spectrum of the BLG double quantum dot is shown in Fig.~\ref{fig_6}(a).
For a single carrier, the system forms a charge qubit governed by the interdot tunnel coupling $t$. The higher-lying Kramers pair, separated by the spin–orbit gap $\Delta_{\mathrm{SO}}$, similarly forms a pair of excited molecular states with the same coupling $t$.
For two carriers in the $(1,1)$ configuration, we consider only the ground-state manifold, which is effectively fourfold degenerate. Three Kramers triplet states~\cite{Triplets_explisit_form} are split from a Kramers singlet by the small exchange energy $J \sim 4t^{2}/U$, with $J \ll k_{\mathrm{B}}T \ll eV_{\mathrm{sd}}$, where $U$ denotes the charging energy~\cite{RevModPhys.79.1217}.

We investigate the DQD in a strongly asymmetric configuration, $\Gamma_{\mathrm{L}}\ll\Gamma_{\mathrm{R}}$ and apply a finite source--drain bias $V_\textnormal{SD}$.
The tunnel couplings between the leads and the individual dots are fixed by the size of the band gap imposed by the inverted dot geometry (see Appendix~\ref{app:fab} for device details), in which the leads are p-doped while the DQD hosts electrons.
To apply the framework developed for a single dot to the DQD, we require that the internal double-dot dynamics, governed by the interdot tunnel coupling $t$, be much faster than tunnelling to the leads, such that $\Gamma_{\mathrm{L}}\ll\Gamma_{\mathrm{R}}\ll t\ll eV_{\mathrm{sd}}$.
This condition ensures that an electron undergoes many interdot transitions before escaping the double dot, effectively mimicking a single quantum dot. 



Accordingly, our rate-equation model assumes incoherent tunnelling into the bonding and antibonding molecular eigenstates, with effective tunnel rates weighted by the corresponding wavefunction amplitudes on each dot as a function of detuning (see Appendix~\ref{app:dqd} for details).
The free parameters of the model are the tunnelling rates $\Gamma_{\mathrm{L/R}}$, the interdot tunnel coupling $t$, and the degeneracy ratio between the initial and final charge states. 
We further assume that inelastic interdot tunnelling (relaxation) is negligible, consistent with its absence in the experimental data.

The simulated occupation and transport current qualitatively reproduce the experimental data in Fig.~\ref{fig_6}(b,c), where charge stability diagrams for both bias polarities are shown around the $(0,0)$–$(1,0)$–$(0,1)$ and $(1,1)$–$(1,0)$–$(0,1)$ triple points. 
We first note that the measured current in Fig.~\ref{fig_6}(c) appears only along a narrow line of width $\sim t$ centered at zero detuning and is blocked elsewhere, indicating that inelastic electron–phonon processes inside the bias triangles are indeed strongly suppressed~\cite{fujisawa1998spontaneous}.

By measuring the full-width at half maximum of the current (or sensor) peak around zero detuning at two opposite bias polarities as shown in Fig.~\ref{fig_6}(b--c):
\begin{equation}
\begin{aligned}
w_{+} &= \frac{t}{\sqrt{2}}\sqrt{\frac{d_{0}\Gamma_{\mathrm{R}}}{d_{1}\Gamma_{\mathrm{L}}}+1},
\\
w_{-} &= \frac{t}{\sqrt{2}}\sqrt{\frac{d_{0}\Gamma_{\mathrm{L}}}{d_{1}\Gamma_{\mathrm{R}}}+1},
\label{eq17}
\end{aligned}
\end{equation}
and the maximum absolute values of the current at zero detuning 
\begin{equation}
\begin{aligned}
\max|I_{+}| &=
\frac{e d_{0}\Gamma_{\mathrm{R}}}{2}
\left(1+\frac{d_{0}}{2d_{1}}\frac{\Gamma_{\mathrm{R}}}{\Gamma_{\mathrm{L}}}\right)^{-1}, \\
\max|I_{-}| &=
\frac{e\,2d_{1}\Gamma_{\mathrm{R}}}{2}
\left(1+\frac{2d_{1}}{d_{0}}\frac{\Gamma_{\mathrm{R}}}{\Gamma_{\mathrm{L}}}\right)^{-1},
\label{eq18}
\end{aligned}
\end{equation}
we obtain sufficient experimental observables to extract an interdot tunnel coupling of $t \approx \SI{2}{\giga\hertz}$ and a tunnelling-rate asymmetry of $\Gamma_{\mathrm{L}}/\Gamma_{\mathrm{R}}\approx 25$, as well as the remaining unknown parameters (see Appendix~\ref{app:dqd}). For clarity, however, we extract the degeneracy using the same approach as for the single quantum dot in the previous section.

From the absolute magnitude of the current at zero detuning, $\max|I_{\pm}| \simeq \SI{5}{\pico\ampere}$, we estimate the slowest tunnelling rate to be $\Gamma_{\mathrm{R}} \sim \SI{30}{\mega\hertz}$. 
This places the device firmly in the regime $\Gamma_{\mathrm{R}}\ll \Gamma_{\mathrm{L}}< t\sim k_{\mathrm{B}}T \ll eV_{\mathrm{bias}}$, very close to the limit in which the asymptotic single-dot approach using Eqs.~\eqref{eq14} and~\eqref{eq15} applies.
We emphasize that a quantitative match between simulation and experiment is not required to extract the degeneracy, because our analysis relies solely on dimensionless ratios derived from Eqs.~\eqref{eq17} and~\eqref{eq18} (see Appendix~\ref{app:dqd} for details).

Figure~\ref{fig_6}(d) shows the finite-bias analysis of the transport current for the electron transition from $0$ to the molecular states $\bigl[(0,1)\pm(1,0)\bigr]/\sqrt{2}$, analogous to the single-dot $0$ to $1$ case in Fig.~\ref{fig_5}(c-–f). 
We extract narrow slices of the charge-stability diagram around zero detuning from Fig.~\ref{fig_6}(c) (white dashed line) at different bias voltages and compile them vertically to emulate a single dot ‘Coulomb diamond’ representation of the bias window in Fig.~\ref{fig_6}(d,i–ii).
The corresponding one-dimensional line cuts along the energy axis, obtained by taking the maximum current across the detuning axis at each energy, are shown in Fig.~\ref{fig_6}(d,iii–iv) for different bias voltages and polarities. 
The pronounced bias asymmetry in the current plateaus demonstrates that the degeneracy of the $\bigl[(0,1)\pm(1,0)\bigr]/\sqrt{2}$ states is nontrivial. 
As in the single-dot case, the ratio of the plateau heights approaches close to the expected value of $d_{{0,1\pm1,0}}=2d_{1}\simeq4$ for $|V_{\mathrm{sd}}|<\Delta_{\mathrm{SO}}$, corresponding to tunnelling into a single Kramers pair whose orbital bonding-antibonding hybridization doubles the degeneracy. The current plateau increases toward $d_{0,1\pm1,0}\simeq8$ at higher biases, where both Kramers doublets contribute equally.

A similar analysis of the $\bigl[(0,1)\pm(1,0)\bigr]/\sqrt{2}\rightarrow(1,1)$ transition in Fig.~\ref{fig_6}(e) shows, in contrast, that the currents at opposite bias polarities are nearly identical within the charge-noise level. 
This indicates the trivial degeneracy ratio of $d_{1,1}/d_{0,1\pm1,0}\simeq1$, as expected for a transition between the fourfold-degenerate one-electron ground state and the fourfold-degenerate $(1,1)$ two-electron ground state depicted in Fig.~\ref{fig_6}(a). 
We do not observe signatures of the anticipated higher excited states of the $(1,1)$ configuration~\cite{ruckriegel2025microwave} entering the bias window at higher $|V_{\mathrm{sd}}|$, likely due to charge noise.

The subtle peak near the edge of the current plateaus, observed in both the simulations and the experimental data in Fig.~\ref{fig_6}(d,e)(iii,iiii), originates from the tunnel-coupling splitting $t$, which is not fully smeared out by the finite electron temperature.
A detailed exploration of the two- and many-electron spectrum in BLG double quantum dots is beyond the scope of the present work, as it would require a measurement setup optimized for low-current detection, along with additional barrier gates to enable the charge-sensor response to become informative.

\section{Conclusions}

The simplicity and universality of our approach provide direct access to degeneracy, and hence entropy of quantum dot systems, in regimes previously inaccessible with existing techniques.
By applying the method to weakly coupled BLG and GaAs quantum dots, we recover the known degeneracy structure in both materials, demonstrating that the technique is robust and platform independent.
This suggests that degeneracy and entropy, long considered thermodynamic equilibrium quantities, can be accessed using standard transport measurements in mesoscopic devices.

More broadly, the non-equilibrium framework can be extended to a wide range of hybrid systems, including quantum dots coupled to fractional quantum Hall states~\cite{PhysRevLett.131.016601}, Kondo clouds~\cite{PhysRevLett.129.227702}, Anderson impurities~\cite{PhysRevB.111.115117}, Majorana modes in continuous systems~\cite{PhysRevLett.123.147702}, as well as in Kitaev quantum-dot chains~\cite{dvir2023realization,tenHaaf2024twosite}, while offering simpler device architectures, faster measurements, and intrinsic energy resolution.

Finally, the data required to extract degeneracies have been routinely collected in standard transport experiments for decades, suggesting that previously overlooked entropy signatures may be recovered across a wide range of quantum-dot systems.

\section{Acknowledgments}
The authors thank Vadim Khrapai and Yigal Meir for interesting discussions and comments. All authors acknowledge financial support by the European Graphene Flagship Core3 Project, H2020 European Research Council (ERC) Synergy Grant under Grant Agreement 951541, the European Innovation Council under grant agreement
number 101046231/FantastiCOF, NCCR QSIT (Swiss National Science Foundation, grant number 51NF40-
185902). K.W. and T.T. acknowledge support from the JSPS KAKENHI (Grant Numbers 21H05233 and
23H02052) and World Premier International Research Center Initiative (WPI), MEXT, Japan.

\section{Corresponding authors}
Correspondence and requests for materials should be addressed to A.D.

\section{Author contributions}
T.T. and K.W. grew the hBN. H.D. fabricated the BLG device.
A.H fabricated the GaAs device.
A.D. designed the experiment, acquired and analyzed the data with inputs from J.R., C.A., T.I. and K.E.. 
J.R. and Z.C. wrote the simulation code.
A.D. wrote the manuscript with inputs from all authors.

\section{Competing interests}
The authors declare no competing financial interests.

\section{Data availability}

The data that support the findings of this study will be made available online through the ETH Research Collection. This includes raw data, analysis scripts, and plotting scripts for figures from the main text.

\bibliography{bibliography_Kramers}

\appendix

\section{GaAs device}
\label{app:gaas}

\begin{figure}[tbh!]
	\includegraphics[width=1\columnwidth]{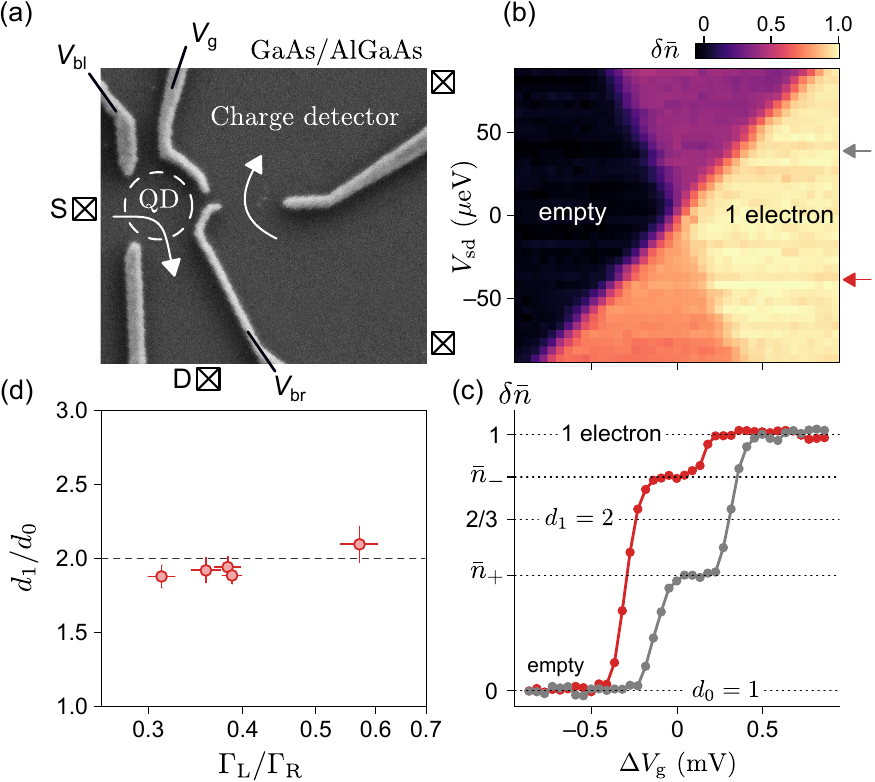}
	\caption{
(a)~Scanning electron microscope image of the GaAs device. The biased single quantum dot is coupled to a quantum point contact charge sensor on the right.
(b)~Measured charge occupation of the dot around the $0$–$1$ electron transition as a function of bias voltage and plunger gate voltage. Colored arrows indicate the positions of the line cuts shown in (c) for negative (red) and positive (gray) bias.
(c)~Charge-occupation line cuts at fixed bias voltages as a function of the plunger gate voltage. Dotted lines mark the occupation plateaus inside the bias window. Symmetric coupling regime plateau of $2/3$ corresponding to a twofold-degenerate spin-$1/2$ state is indicated.
(d)~Extracted degeneracy of the single-electron state in the GaAs quantum dot as a function of dot–lead coupling asymmetry. The data lie close to the expected value of $2$.
}
	\label{fig_8}
\end{figure}

To demonstrate the applicability of our technique to other material platforms, we extract the degeneracy of a single electron in a GaAs quantum dot, shown in Fig.~\ref{fig_8}(a). To do so, we analyze an earlier data set originally acquired for a different purpose, thereby illustrating that degeneracy information in quantum-dot transport measurements has often remained unrecognized.
The charge-sensor signal around the $0-1$ electron transition is shown as a function of plunger gate voltage and bias voltage in Fig.~\ref{fig_8}(b).
Similar to bilayer graphene quantum dots, the relatively flat occupation plateaus inside the bias window indicate that the bias voltage in GaAs does not significantly modify the tunnel-barrier couplings to the dot.
The colored arrows indicate the positions of the line cuts taken at opposite bias polarities and shown in Fig.~\ref{fig_8}(c).

The device operates in the intermediate coupling regime, $\Gamma_{\mathrm{L}}/\Gamma_{\mathrm{R}} \sim \numrange{0.3}{0.6}$, for which a bias asymmetry of the occupation plateaus around the symmetric value $\delta\bar{n}=2/3$ is expected.
The extracted single-electron degeneracy obtained using Eq.~\eqref{eq11} is shown in Fig.~\ref{fig_8}(c) as a function of the coupling asymmetry.
The classical result $d_{1}=2$ for a spin-degenerate GaAs quantum dot~\cite{hartman2018direct} is clearly reproduced.

\section{General rate-equation framework}
\label{app:framework}

Two measured quantities, the dot occupation and the transport current, are determined~\cite{ihn2010semiconductor} by the steady-state probabilities of the microstates for both manifolds $\{p^{\mathrm{n}}_{i}\}$ and $\{p^{\mathrm{n+1}}_{j}\}$ as well as by forward $W^{+}_{ij}=W^{+L}_{ij}+W^{+R}_{ij}$ and reverse $W^{-}_{ji}=W^{-L}_{ji}+W^{-R}_{ji}$ transition rates as indicated by the blue arrows in Fig.~\ref{fig_1}(c):

\begin{align}
\delta \bar{n} &= \sum^{d_{\mathrm{n+1}}}_{j=1} p^{\mathrm{n+1}}_{j}, \label{eq1} \\
I &= e \sum^{d_{\mathrm{n}}}_{i=1} \sum^{d_{\mathrm{n+1}}}_{j=1} \Big[ p^{\mathrm{n}}_{i} W^{+R}_{ij} - p^{\mathrm{n+1}}_{j} W^{-R}_{ji} \Big]\label{eq2}. 
\end{align}

Transitions are induced by tunnelling from the left and right leads, where carriers of all possible flavors are available and thermalized with base electron temperature.
The unknown set of $\{d_{\mathrm{n}} + d_{\mathrm{n+1}}\}$ probabilities is determined by a corresponding set of $\{d_{\mathrm{n}} + d_{\mathrm{n+1}}\}$ steady-state balance equations together with the normalization condition:

\begin{equation}
\left\{
\begin{aligned}
&-p^{\mathrm{n}}_{i}\sum^{d_{\mathrm{n+1}}}_{j=1}W^{+}_{ij}+\sum^{d_{\mathrm{n+1}}}_{j=1} p^{\mathrm{n+1}}_{{j}}W^{-}_{ji}=0,~i = 1 \dots d_{\mathrm{n}}, \\
&-p^{\mathrm{n+1}}_{j}\sum^{d_{\mathrm{n}}}_{i=1}W^{-}_{ji}+\sum^{d_{\mathrm{n}}}_{i=1} p^{\mathrm{n}}_{{i}}W^{+}_{ij}=0,~j = 1 \dots d_{\mathrm{n+1}}, \\
&\sum^{d_{\mathrm{n}}}_{i=1} p^{\mathrm{n}}_{{i}}+\sum^{d_{\mathrm{n+1}}}_{j=1} p^{\mathrm{n+1}}_{{j}}=1
\end{aligned}
\right.
\label{eq3}
\end{equation}

In the case of an irreducible graph, where the states separate into two degenerate groups, the much more constrained set of detailed balance equations $\Big[p^{\mathrm{n}}_i W^{+}_{ij} = p^{\mathrm{n+1}}_{j} W^{-}_{ji},~\forall i,~j\Big]$ already has a consistent solution, which instantly solves the steady-state balance from~\eqref{eq3}. 

Indeed, according to the Fermi's golden rule, the ratio of the forward and backward transition rates $\kappa={W^{+}_{ij}}/{W^{-}_{ji}}$ is independent of the specific indices $i$ and $j$ chosen within the degenerate manifolds (or, more generally, for any pair of states whose transition energy lies well inside the bias window):

\begin{widetext}
\begin{align}
\kappa=\frac{W^{+}_{ij}}{W^{-}_{ji}}
=
\frac{
\cancel{\frac{2\pi}{\hbar}\,\rho(\Delta E_{ij})
 \sum_{\lambda}\left|\langle j| a_{\lambda}^{\dagger} | i\rangle \right|^{2}}
 \big[ \Gamma_{\mathrm{L}} f_{\mathrm{L}}(\Delta E_{ij})
   + \Gamma_{\mathrm{R}} f_{\mathrm{R}}(\Delta E_{ij}) \big]
}{
\cancel{\frac{2\pi}{\hbar}\,\rho(\Delta E_{ij})
 \sum_{\lambda}\left|\langle i| a_{\lambda} | j\rangle \right|^{2}}
 \big[ \Gamma_{\mathrm{L}} (1 - f_{\mathrm{L}}(\Delta E_{ij}))
   + \Gamma_{\mathrm{R}} (1 - f_{\mathrm{R}}(\Delta E_{ij})) \big]
} 
= \frac{f^{*}(\Delta E_{ij})}{1 - f^{*}(\Delta E_{ij})},
\label{eq4}
\end{align}
\end{widetext}

where, $\Delta E_{ij} = E_j - E_i$ denotes the chemical potential of the quantum dot, $\rho$ is the density of states in the leads, $a^{\dagger}_\lambda$ ($a_\lambda$) adds (removes) a carrier in the single-particle state $\lambda$ to (from) the dot, and $f_{\mathrm{L}}(\Delta E_{ij})$ and $f_{\mathrm{R}}(\Delta E_{ij})$ are the equilibrium Fermi--Dirac distributions in the leads. The effective distribution function seen by the dot,

\begin{equation}  
f^{*}=\frac{\Gamma_{\mathrm{L}}}{\Gamma_{\mathrm{L}}+\Gamma_{\mathrm{R}}}f_{\mathrm{L}}+\frac{\Gamma_{\mathrm{R}}}{\Gamma_{\mathrm{L}}+\Gamma_{\mathrm{R}}}f_{\mathrm{R}}
\label{eq5}
\end{equation},

is a non-equilibrium linear combination (double-step) of the thermal distributions in the leads~\cite{PhysRevLett.79.3490, PhysRevLett.102.036804,Altimiras2010, PhysRevB.102.085417}. 
It is important to emphasize that the factorization of the matrix elements in Eq.~\eqref{eq4} for both the left and right barriers is valid only if, for each given state, the orbital part of the wavefunction has identical overlap with the left and right reservoirs, even though the wavefunctions of different degenerate states may differ.
This condition can be realized, for example, if the quantum dot has an approximate elliptical symmetry with respect to the left and right tunnel barriers or, ideally, if the two reservoirs are brought sufficiently close together such that they probe the same wavefunction tail of the dot.

Due to the proportionality from Eq.~\eqref{eq4}, the resulting Markov chain satisfies the Kolmogorov criterion for time reversibility~\cite{kelly1979reversibility}: for any closed loop of states, the product of the forward transition probabilities equals that of the corresponding time-reversed loop. For example,

\begin{equation}
\frac{W^{+}_{ij} W^{-}_{jk} W^{+}_{kl} W^{-}_{li}}{W^{+}_{il} W^{-}_{lk} W^{+}_{kj} W^{-}_{ji}}
= \kappa \frac{1}{\kappa} \, \kappa \frac{1}{\kappa} = 1
\label{eq6}
\end{equation}

for any four connected states $i,j,k,l$. 
This guarantees that detailed balance holds and that all probabilities within the corresponding manifolds become identical, $p^{\mathrm{n+1}}_{j} = p^{\mathrm{n}}_{i} W^{+}_{ij}/W^{-}_{ji} = \kappa\, p^{\mathrm{n}}_{i}$ for any $i$ and $j$.
Therefore, the average occupation of the dot at large positive (negative) bias, $|eV_{\mathrm{sd}}| \gg k_{\mathrm{B}}T$ (where $f_{\mathrm{L}} \to 1\,(0)$ and $f_{\mathrm{R}} \to 0\,(1)$), as well as the ratio of the positive and negative currents, depends only on two dimensionless parameters: the ratio of degeneracies between the initial and final states ${d_{\mathrm{n+1}}}/{d_{\mathrm{n}}}$ and the asymmetry between tunnelling rates $\Gamma_{\mathrm{L}}/\Gamma_{\mathrm{R}}$.

\section{Beyond weak coupling}
\label{app:strongc}

\begin{figure}[tbh!]
	\includegraphics[width=1\columnwidth]{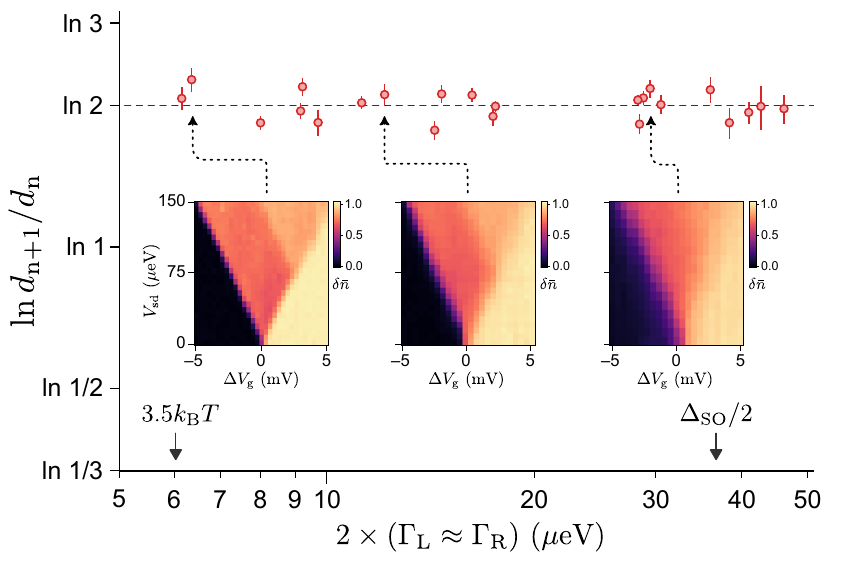}
	\caption{Extracted ground-state degeneracy of the first electron in a bilayer graphene quantum dot as a function of the dot–lead coupling rate in the nearly symmetric regime $\Gamma_{\mathrm{L}}\approx\Gamma_{\mathrm{R}}$. Insets show the measured dot occupation as a function of bias voltage and plunger gate voltage for the corresponding tunnelling rates indicated by the arrows. The highest coupling rate reaches approximately half of the spin–orbit gap, for which the occupation plateau remains clearly resolved.}
	\label{fig_7}
\end{figure}

We verify that the absolute strength of the dot–lead tunnel coupling, $\Gamma_{\mathrm{L}}+\Gamma_{\mathrm{R}}$, in the nearly symmetric regime ($\Gamma_{\mathrm{L}}\approx\Gamma_{\mathrm{R}}$) does not affect the relative height of the occupation plateau inside the bias window, and therefore yields the same extracted degeneracy.
Figure~\ref{fig_7} shows the extracted ground-state degeneracy (entropy) for the first electron in a bilayer graphene quantum dot for different overall coupling strengths.
The insets display finite-bias measurements across the $0-1$ transition for the corresponding couplings, indicated by arrows.

Over a wide range of tunnelling rates, from values below the thermal broadening scale $3.5 k_{\mathrm{B}}T \approx \SI{6}{\micro\electronvolt}$ up to approximately half the spin–orbit gap, $\Delta_{\mathrm{SO}}/2 \approx \SI{37}{\micro\electronvolt}$, the extracted degeneracy remains consistent with the expected value $d_{1}=2$.
Further increasing the dot–lead coupling causes the occupation plateaus to broaden and overlap with the excited-state plateau, making the ground-state plateau hardly resolvable, as illustrated in the insets in Fig.~\ref{fig_7}.

\section{Fabrication and measurement}
\label{app:fab}

The BLG device fabricated on a van der Waals heterostructure stacked by standard dry transfer techniques. The stack consists of a \SI{40}{nm} thick top hBN layer, the Bernal BLG sheet, a \SI{48}{nm} thick bottom hBN, and a graphite back gate layer. The ohmic contacts to the BLG are 1D edge contacts. To form QDs, we utilize the two overlapping layers of Cr/Au (\SI{3}{nm}/\SI{20}{nm}) metallic gates. The upper gate layer consists of finger gates which are deposited on top of a \SI{26}{nm} thick insulating aluminum oxide layer. 

\begin{figure}[tbh!]
	\includegraphics[width=0.95\columnwidth]{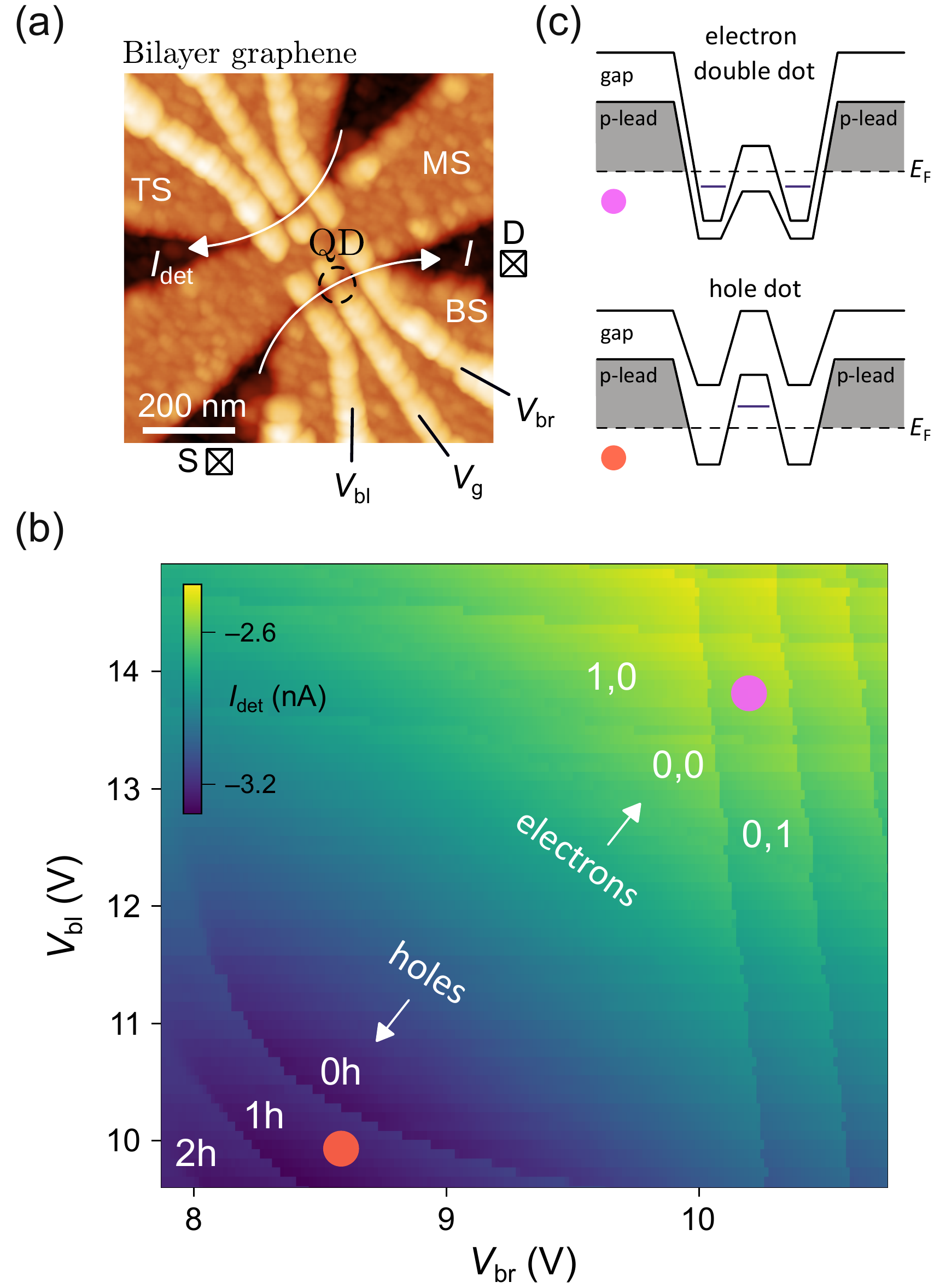}
	\caption{
(a)~Atomic force microscope image of the bilayer graphene device. Top (TS), middle (MS), and bottom (BS) split gates define the conducting channels, while finger gates locally confine carriers to form quantum dots. The global back gate is held at a constant negative voltage $V_{\mathrm{bg}} = \SI{-6}{\volt}$, thereby p-doping the leads.
(b)~Charge-stability diagram of a bilayer graphene quantum dot defined by three finger gates. The carrier type is controlled by the barrier gate voltages, which shift the conduction-band edge below (electrons) or above (holes) the Fermi level in the leads.
(c)~Schematic energy diagram illustrating the formation of electron and hole quantum dots for p-doped leads. Colored circles indicate the corresponding operating points in the charge-stability diagram map shown in (b).
}
	\label{fig_9}
\end{figure}

A global back gate voltage of $V_{\mathrm{BG}}=\SI{-6}{V}$ p-dopes the BLG and, together with metallic split gates, is capable of inducing a displacement field up to $D=\SI{-0.46}{V/nm}$. 
The lower metallic gate layer opens a band gap of order $\SI{50}{meV}$ beneath the top (TS), middle (MS), and bottom (BS) split gates, thereby defining two conducting channels, as shown in Fig.~\ref{fig_9}(a). 
Quantum dots are formed inside these channels using an upper layer of narrow finger gates.

The tunnel couplings to the left and right reservoirs are controlled by the barrier gate voltages $V_{\mathrm{bl}}$ and $V_{\mathrm{br}}$, while the plunger gate voltage $V_{\mathrm{g}}$ tunes the dot into the single-electron or single-hole regime. 
The charge-sensor stability diagram, measured as a function of the barrier gate voltages, is shown in Fig.~\ref{fig_9}(b).

When the lead Fermi level lies above the dot chemical potentials under the barrier gates and within the band gap under the middle gate, as shown in Fig.~\ref{fig_9}(c), two electron dots are formed, producing the honeycomb pattern visible in Fig.~\ref{fig_9}(b). 
In this configuration, the dot–lead tunnel couplings are mostly fixed by the band gap, while the interdot coupling is highly tunable via the middle gate. 
When the lead Fermi level lies within the band gap under the barrier gates but above the dot chemical potential under the plunger gate, as shown in Fig.~\ref{fig_9}(c), a single hole dot is formed, with tunnel couplings that are highly tunable by the barrier gates.

The GaAs quantum dot device is fabricated in a GaAs/AlGaAs heterostructure hosting a two-dimensional electron gas located $\SI{90}{\nano\metre}$ below the surface. 
The electron density and mobility are $n = \SI{2e11}{\per\centi\metre\squared}$ and $\mu = \SI{2e6}{\centi\metre\squared\per\volt\second}$, respectively, as measured at $\SI{1.3}{\kelvin}$.

Metallic top gates (shown in gray in Fig.~\ref{fig_8}(a)) are used to electrostatically confine electrons in the two-dimensional electron gas by applying negative voltages. 
The quantum dot is tuned to the single-electron regime. 
Barrier gate voltages $V_{\mathrm{bl}}$ and $V_{\mathrm{br}}$ control the tunnel rates to the left and right leads, respectively, while the plunger gate voltage $V_{\mathrm{g}}$ sets the electron number in the dot. 
An additional gate on the left side of the dot defines a quantum point contact, which serves as a charge sensor to measure the average dot occupation.

Samples are mounted on the mixing chamber of a Oxford dilution refrigerator, which has a base temperature of $\SI{11}{\milli\kelvin}$ and an independently extracted electron temperature of around $\SIrange{20}{30}{\milli\kelvin}$, depending on the state of the device. All the measurement and control electronics are located at room temperature and are connected to the device via 24 DC lines. Each line is low-pass filtered using RC filters mounted on the mixing chamber plate, with a time constant of approximately $\SI{10}{\micro\second}$. For DC biasing of gates and ohmic contacts, we use in-house built low-noise voltage sources with a cutoff frequency of $\SI{7}{\hertz}$.

\section{Double Quantum Dot Model}
\label{app:dqd}
\begin{figure*}[tbh!]
	\includegraphics[width=2\columnwidth]{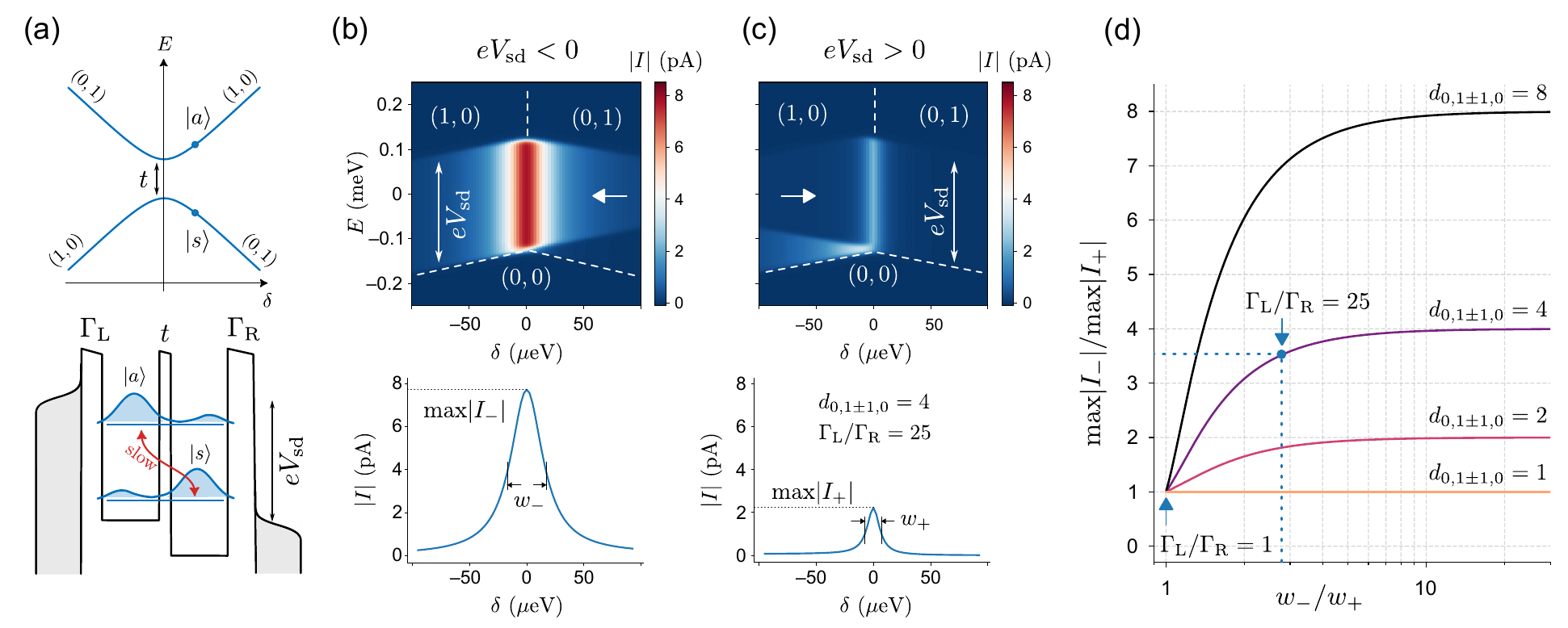}
	\caption{(a)~Upper panel: Energy diagram of a single charge carrier in a double quantum dot with interdot tunnel coupling $t$. The two eigenstates are denoted $|a\rangle$ and $|s\rangle$. 
    Lower panel: Energy diagram of the double quantum dot tunnel-coupled to the leads at finite detuning $\delta\neq 0$. Relaxation and excitation processes between the two eigenstates (red arrow) are assumed to be much slower than the tunnelling rates.
    (b)~Upper panel: Rate-equation simulation of the absolute value of the transport current through the double quantum dot as a function of detuning and energy offset for positive and negative bias voltages $V_{\mathrm{sd}} = \SI{\pm250}{\micro\electronvolt}$. 
    Lower panel: Horizontal line cuts taken from the upper panel at the positions indicated by the white arrows. The simulations correspond to a fourfold-degenerate Kramers ground-state manifold of the double dot, $d_{0,1\pm1,0}=2\times2=4$, in a highly asymmetric dot lead coupling regime with $\Gamma_{\mathrm{L}}/\Gamma_{\mathrm{R}}=25$.
    (d)~Ratios of the maximum absolute current values, $\max|I_{\pm}|$, at zero detuning $\delta=0$, and of the corresponding full widths at half maximum, $w_{\pm}$, shown for different total degeneracies of the single-electron state $d_{0,1\pm1,0}$. The blue point corresponds to the case shown in (b).
    }
	\label{fig_10}
\end{figure*}

We employ a rate-equation approach to calculate the transport current and the charge occupation of the double quantum dot shown in Fig.~\ref{fig_10}(a) as functions of the detuning $\delta$ and the energy offset $E$.
We consider the regime in which the interdot tunnel coupling is much faster than tunnelling to the leads, $t \gg \Gamma_{\mathrm{L,R}}$, while remaining smaller than the thermal energy and much smaller than the applied bias voltage,
\begin{equation}
\Gamma_{\mathrm{L,R}}\ll t < k_{\mathrm{B}}T \ll eV_{\mathrm{sd}}.
\end{equation}
We further neglect relaxation and excitation processes between the predominantly symmetric $|s\rangle$ and predominantly antisymmetric $|a\rangle$ eigenstates of the DQD, assuming their characteristic times to be much longer than the slowest tunnelling rate.

Under these assumptions, the electron wavefunction, after tunnelling into, for example, the left dot and before escaping to the right lead, is not localized in the left dot but rapidly becomes distributed according to the weights of the corresponding molecular eigenstates at a given detuning~\cite{RevModPhys.75.1}.

\begin{equation}
\frac{|\langle s|R\rangle|^2}{|\langle s|L\rangle|^2}
=
\frac{|\langle a|L\rangle|^2}{|\langle a|R\rangle|^2}
=
\frac{t^2/4}{\bigl(\sqrt{\delta^2+t^2/4}-\delta\bigr)^2}.
\end{equation}

In other words, tunnelling from the leads is treated as incoherent and occurs directly into one of the molecular eigenstates, without subsequent relaxation.
At detunings much larger than the tunnel coupling $|\delta|\gg t$, where the wavefunction becomes strongly localized on one dot, the electron is effectively trapped in the dot into which it initially tunnels.

For clarity, here we analytically solve the three-state rate-equation problem in the limit where both molecular eigenstates lie deep inside the bias window~$|eV_{\mathrm{sd}}|\gg k_{\mathrm{B}}T,|\delta|$. In this regime, the electron temperature is not a relevant energy scale and the distribution functions in the leads reduce to $f_{\mathrm{L}}=1$ and $f_{\mathrm{R}}=0$. 
All numerical simulations presented in the main text and appendix nevertheless fully account for finite temperature effects.

The steady-state probabilities for the DQD to be empty, or occupied in the states $|s\rangle$ and $|a\rangle$, obey master equation:
\begin{equation}
\left\{
\begin{aligned}
&\dot{p}_{\mathrm{s}}
=-p_{\mathrm{s}}\, d_{0} |\langle s|R\rangle|^2 \Gamma_{\mathrm{R}}
+ p_{0}\, d_{1} |\langle s|L\rangle|^2 \Gamma_{\mathrm{L}} = 0, \\
&\dot{p}_{\mathrm{a}}
=-p_{\mathrm{a}}\, d_{0} |\langle a|R\rangle|^2 \Gamma_{\mathrm{R}}
+ p_{0}\, d_{1} |\langle a|L\rangle|^2 \Gamma_{\mathrm{L}} = 0, \\
& p_{\mathrm{s}} + p_{\mathrm{a}} + p_{0} = 1 ,
\end{aligned}
\right.
\end{equation}
where the tunnelling rates are rescaled by the corresponding wavefunction weights on each dot, and $d_{0}$ and $d_{1}$ denote the degeneracies of the empty and singly occupied manifolds, respectively. The resulting absolute value of the transport current through the DQD is therefore bias dependent, yielding Lorentzian peaks
\begin{equation}
\begin{aligned}
|I_{+}| &= e\,\frac{d_{1}\Gamma_{\mathrm{L}}\, d_{0}\Gamma_{\mathrm{R}}}
{d_{0}\Gamma_{\mathrm{R}} + d_{1}\bigl(2+16\delta^2/t^2\bigr)\Gamma_{\mathrm{L}}}, \\
|I_{-}| &= e\,\frac{d_{1}\Gamma_{\mathrm{R}}\, d_{0}\Gamma_{\mathrm{L}}}
{d_{0}\Gamma_{\mathrm{L}} + d_{1}\bigl(2+16\delta^2/t^2\bigr)\Gamma_{\mathrm{R}}}.
\end{aligned}
\label{app_eq_Ipm}
\end{equation}
for positive ($+$) and negative ($-$) polarities.

Figure~\ref{fig_10}(b) shows the finite-temperature solutions for two opposite bias voltages, $V_{\mathrm{sd}} = \SI{\pm250}{\micro\electronvolt}
 \gg k_{\mathrm{B}}T \simeq \SI{3}{\micro\electronvolt}$, in the case of asymmetric dot–lead coupling $\Gamma_{\mathrm{L}}/\Gamma_{\mathrm{R}}=25$.
We consider the experimentally relevant case discussed in the main text: a bilayer graphene DQD with a Kramers-doublet degeneracy $d_{1}=2$ for the single-carrier state and a trivial degeneracy $d_{0}=1$ for the empty DQD.
At zero detuning, the nearly degenerate molecular states $(|10\rangle\pm|01\rangle)/\sqrt{2}$ introduce an additional orbital degree of freedom, such that the total degeneracy becomes $d_{0,1\pm1,0}=2d_{1}=4$.

The horizontal Lorentzian line cuts taken deep inside the bias window and shown in the lower panels of Fig.~\ref{fig_10}(b) are described by Eq.~\eqref{app_eq_Ipm} as functions of the detuning.
The full widths at half maximum of the current peaks are set by the interdot tunnel coupling and the dot–lead coupling asymmetry,
\begin{align}
w_{+} &= \frac{t}{\sqrt{2}}\sqrt{\frac{d_{0}\Gamma_{\mathrm{R}}}{d_{1}\Gamma_{\mathrm{L}}}+1},
\label{aeq16}\\
w_{-} &= \frac{t}{\sqrt{2}}\sqrt{\frac{d_{0}\Gamma_{\mathrm{L}}}{d_{1}\Gamma_{\mathrm{R}}}+1},
\label{aeq17}
\end{align}
from which both parameters can be extracted independently.

The maximum absolute values of the current at zero detuning are given by
\begin{equation}
\begin{aligned}
\max|I_{+}| &=
\frac{e d_{0}\Gamma_{\mathrm{R}}}{2}
\left(1+\frac{d_{0}}{2d_{1}}\frac{\Gamma_{\mathrm{R}}}{\Gamma_{\mathrm{L}}}\right)^{-1}, \\
\max|I_{-}| &=
\frac{e\,2d_{1}\Gamma_{\mathrm{R}}}{2}
\left(1+\frac{2d_{1}}{d_{0}}\frac{\Gamma_{\mathrm{R}}}{\Gamma_{\mathrm{L}}}\right)^{-1},
\end{aligned}
\end{equation}
where the factor of two in front of $d_{1}$ arises from the orbital degeneracy.

The ratios of the peak currents together with the ratios of the corresponding widths provide two experimentally accessible, dimensionless quantities that depend on the two unknown parameters $d_{1}/d_{0}$ and $\Gamma_{\mathrm{L}}/\Gamma_{\mathrm{R}}$.
This allows the coupling asymmetry to be eliminated and the degeneracy ratio to be extracted directly.
The resulting relationship is shown in Fig.~\ref{fig_10}(c), where for each $d_{0,1\pm1,0}=2d_{1}$ we plot $\max|I_{-}|/\max|I_{+}|$ as a function of $w_{-}/w_{+}$.

For symmetric coupling, $\Gamma_{\mathrm{L}}/\Gamma_{\mathrm{R}}=1$, both the widths and heights of the current peaks at opposite bias polarities are identical for any degeneracy.
In the opposite limit, $\Gamma_{\mathrm{L}}/\Gamma_{\mathrm{R}}\gg1$, the current ratio rapidly saturates at the value corresponding to $2d_{1}/d_{0}$, in close analogy with the single dot case.
For example, for $w_{-}/w_{+}\simeq2.7$, corresponding to $\Gamma_{\mathrm{L}}/\Gamma_{\mathrm{R}}=25$ and $2d_{1}/d_{0}=4$, the current ratio already reaches approximately $90\%$ of its asymptotic value.

\end{document}